\documentclass[aps,prx,noshowpacs,twocolumn,superscriptaddress,nobibnotes]{revtex4-1}

\usepackage{natbib}
\usepackage{times}
\usepackage{latexsym,amsmath}
\usepackage{amsmath}
\usepackage{graphicx}
\usepackage{multirow}
\usepackage{epstopdf}
\usepackage{amssymb}
\usepackage{epsfig}
\usepackage{xcolor}
\usepackage[normalem]{ulem}
\usepackage{url}
\usepackage{hyperref}
\usepackage{float}
\usepackage{pdfpages}
\makeatletter
\AtBeginDocument{\let\LS@rot\@undefined}
\makeatother

\begin{document}

\title{Optimal navigability of weighted human brain connectomes in physical space}
\date{\today}
	\author{Laia Barjuan}
	\affiliation{Departament de F\'isica de la Mat\`eria Condensada, Universitat de Barcelona, Mart\'i i Franqu\`es 1, E-08028 Barcelona, Spain}
	\affiliation{Universitat de Barcelona Institute of Complex Systems (UBICS), Universitat de Barcelona, Barcelona, Spain}
	\author{Jordi Soriano}
	\affiliation{Departament de F\'isica de la Mat\`eria Condensada, Universitat de Barcelona, Mart\'i i Franqu\`es 1, E-08028 Barcelona, Spain}
	\affiliation{Universitat de Barcelona Institute of Complex Systems (UBICS), Universitat de Barcelona, Barcelona, Spain}
	\author{M. {\'A}ngeles Serrano}
	\email[]{marian.serrano@ub.edu}
	\affiliation{Departament de F\'isica de la Mat\`eria Condensada, Universitat de Barcelona, Mart\'i i Franqu\`es 1, E-08028 Barcelona, Spain}
	\affiliation{Universitat de Barcelona Institute of Complex Systems (UBICS), Universitat de Barcelona, Barcelona, Spain}
	\affiliation{ICREA, Passeig Llu\'is Companys 23, E-08010 Barcelona, Spain}

\begin{abstract}
The architecture of the human connectome supports efficient communication protocols relying either on distances between brain regions or on the intensities of connections.
However, none of these protocols combines information about the two or reaches full efficiency. Here, we introduce a continuous spectrum of 
decentralized routing strategies that combine link weights and the spatial embedding of connectomes to transmit signals. We applied the protocols to individual connectomes in two cohorts, and to cohort archetypes designed to capture weighted connectivity properties. We found that there is an intermediate region, a {\em sweet spot}, in which navigation achieves maximum communication efficiency at low transmission cost. Interestingly, this phenomenon is robust and independent of the particular configuration of weights. 
Our results indicate that the intensity and topology of neural connections and brain geometry interplay to boost communicability, fundamental to support effective responses to external and internal stimuli and the diversity of brain functions. 
\end{abstract}

\maketitle

\section{Introduction}

The architecture of the human brain has been evolutionary shaped in a 3-dimensional Euclidean space to optimize specialization and adaptation to a changing environment~\cite{stiso2018,bullmore2012economy,ercsey2013predictive,schwartz2023evolution}. As a result, communication processes in the brain are extremely efficient, enabling ultrafast responses to a diversity of external and internal stimuli. At the macroscopic level, the transfer of information in these communication processes is sustained by neural networks of white matter fibers that connect neurons across different brain regions, and whose wiring diagrams are conventionally named connectomes~\cite{sporns2005human}. Human connectomes are complex, and their structural features ---such as small worldness~\cite{watts1998collective,sporns2004organization,bassett2006small, he2007small,hagmann2007mapping,bassett2017small}, heterogeneous degree distribution~\cite{gong2009mapping,gastner2016topology}, rich-club effect~\cite{van2011rich}, and modularity~\cite{sporns2016modular, meunier2010modular}--- influence decisively communication mechanisms and all our cognitive processes~\cite{bressler2010large,deco2014great,honey2007network}. 

Understanding the intricacy of the human connectome and its impact on cognitive processes is a daunting task. However, when its topological organization is combined with its anatomical spatial embedding, the human connectome reveals itself as a comprehensive, highly useful map of the brain to understand large-scale neural communication. In recent years, modeling based on these maps in combination with geometric routing protocols have shed light on the informational cost associated with the selection of efficient routes in the brain~\cite{Seguin2018,Allard2020, cannistraci2022geometrical,seguin2023brain}. 

The most efficient routes would correspond to pathways with short topological path lengths that decrease conduction latency, minimize the impact of noise introduced by synaptic retransmission, and reduce metabolic costs~\cite{AK2018}, but their computation requires full knowledge of the connectome's topology. By contrast, a decentralized greedy routing dynamics on geometric maps of connectomes, guided by a local rule that sends information to the connected region closest in Euclidean distance to a target destination, has shown that brain networks are highly navigable~\cite{Seguin2018, Allard2020}, with an efficiency similar to topological shortest path routing but without requiring information about all possible routes, a condition that is highly unlikely in a physiological system. 

As an alternative to spatial distances, routing strategies can be guided by the link weights (e.g., nerve fiber density) in the connectome once these weights are transformed into weight distances such that higher weights correspond to shorter weight distances. In~\cite{AK2019}, weight distances between connected brain regions were used to explore a continuous spectrum of stochastic protocols that interpolate between random-walk diffusion at one extreme, using only local information but performing inefficiently, and shortest path routing at the other, highly efficient but requiring full knowledge of the global weighted topology of the brain connectome. In between, a small increase in the bias towards global information may progressively achieve improved efficiency at low informational cost. 

In this work, we introduce a theoretical framework to investigate the interplay between spatial distances and link weights in communication processes between brain regions, and apply it to human connectomes reconstructed from empirical data. Our framework facilitates the characterization of the distinctive roles played by the \textit{hard} and \textit{soft} wirings of the brain. The former describes the topology of the connectome and the spatial distances in its geometric embedding, both shaped by evolutionary processes. The latter entails the weights of the links between connected brain regions, affected by plasticity and functional needs. More specifically, we applied a continuous spectrum of stochastic routing protocols, having the greedy routing strategy at one extreme and a weight-biased random walk at the opposite extreme, on two cohorts of real human connectome networks. We found that there is an intermediate region in this spectrum, a {\em sweet spot}, in which connectomes become maximally navigable and achieve full communication efficiency. Additionally, in this region, weights, topology and distances are coupled in such a way that information transmission not only is maximally efficient but also robust even under severe insult. Interestingly, this phenomenon is independent of the particular configuration of weights.

\section{Results} 
\subsection{Empirical data and group representative} 
Our study is based on two different datasets with a total of 84 weighted connectomes of healthy human subjects, previously analyzed in~\cite{Zheng2020} to assess their multiscale organization. In the present work, we used the highest resolution layer, obtained by parcellating brains in 1015 regions of interest (ROIs) with approximately equal surface area. The first dataset (University of Lausanne; UL) contains 40 subjects (16 females). The second dataset (Human Connectome Project; HCP)~\cite{van2012human}, is used to cross-validate the results and contains 44 subjects (31 females). All connectomes encompass both hemispheres and comprise typically $N=1014$ nodes without the brainstem region. See Data description in the Materials and Methods section for more details.

The connection weight between pairs of regions was measured as the fiber density of white matter tracts, i.e., number of streamlines connecting the two brain regions per unit surface, where each streamline is corrected by its average length in millimeters, see Methods. The UL dataset is substantially sparser than the HCP dataset (mean of the average degrees is 27.62 in UL and 78.43 in HCP). This introduces some quantitative differences but, overall, the results are qualitatively similar in the two cohorts. We used the coordinates of the centers of the regions in 3D Euclidean space to compute spatial distances between them, also measured in millimeters. 

We also constructed a group-representative weighted connectome for each dataset to disregard subject-to-subject fluctuations. We defined our group-representative connectome as the network formed by the connections that appeared in at least $30 \%$ of individuals in the dataset. This percentage value was selected such that the group-representative had an average degree as close as possible to the mean average degree of the cohort. In order to assign a weight to one of the connections in the group-representative, we collected the set of weights associated with that connection from all the connectomes of the dataset where it was present, and selected one of them randomly. Different realizations of this selection process gave similar results. One interesting fact about our approach is that the group-representative not only preserves the topological features of the original connectomes but also the weight properties.

\subsection{Correlation between weights and spatial distances in human connectomes} 
\begin{figure}[t!]
    \centering
    \includegraphics[width=0.49\textwidth]{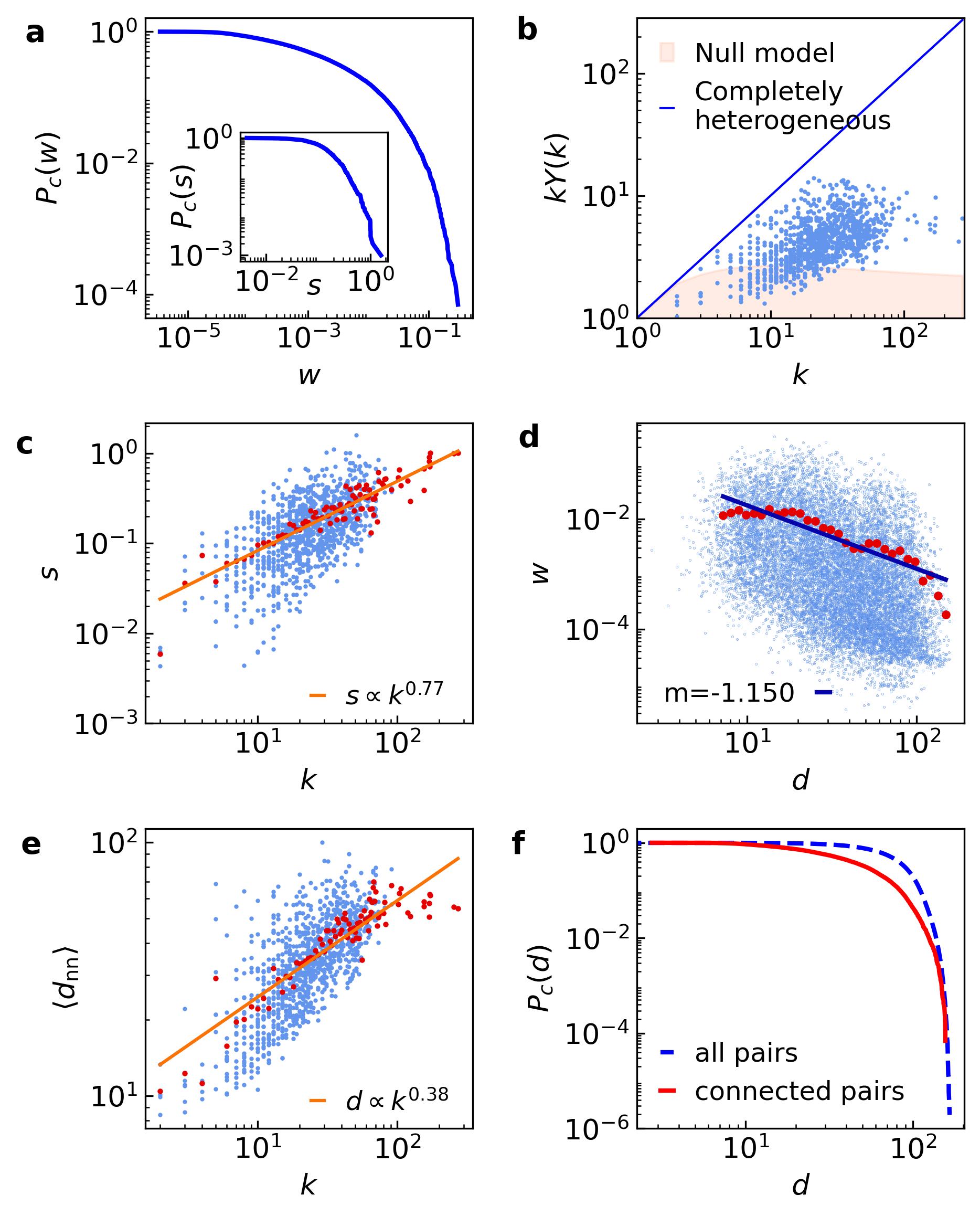}
    \caption{\textbf{Weights and distances in the connectome of subject 34 in the UL dataset. a)} Complementary cumulative weight distribution. Inset: complementary cumulative strength distribution. \textbf{b)} Scattered plot of the disparity of weights, see Methods.  The blue line represents the completely heterogeneous scenario, $kY_i(k) =k$.  The orange area corresponds to the null model presented in \cite{MAS2009}, which describes an intermediate behavior between complete heterogeneity and complete homogeneity,  $kY_i(k) =1$. \textbf{c)} Strength versus degree. The red dots are the average values of strength for each degree. The orange curve is the fit of the averages. \textbf{d)} Weight versus Euclidean distance. The red dots are the average values of weight in the range from $2.5\cdot {\rm d}_{\rm min}$ to ${\rm d}_{\rm max}$. We computed the averages in 30 logarithmically spaced bins. The blue line is the fit of the averages. \textbf{e)} Average distance to nearest neighbors versus degree. The red dots are the average values of distance for each degree and the orange curve is the fit of the averages. \textbf{f)} Complementary cumulative distance distribution. The dashed blue curve represents the distribution of distances between all pairs of regions in the network and the solid red curve between connected pairs.}
    \label{fig:weightdists}
\end{figure}

The properties of connection weights and spatial distances between regions, as well as their interrelation, are shown in Fig.~\ref{fig:weightdists} for a typical subject in the UL dataset. Results for the rest of UL connectomes are reported in Figs.~SF1-SF9, and in Figs.~SF55-SF63 for the connectomes in the HCP dataset. In general, we observed that the results were consistent across subjects.

Figure~\ref{fig:weightdists}a shows the complementary cumulative distribution of weights $P_c (w)$, which ranges over at least four orders of magnitude, and Fig.~\ref{fig:weightdists}b shows the local distribution of weights around each brain ROI as measured by the disparity $\Upsilon_i (k)$, see Methods. These features indicate that the distribution of weights is clearly heterogeneous both at the global and local levels, and show that very different weight scales are present in human brain connectomes. In particular, the disparity measure reveals that specific connections concentrate most of the strength associated to brain regions. The construction of our group-representative is interesting because it maintains the weight distribution and the disparity property, see Figs.~SF17-SF18 and SF70. This is in contrast to the results obtained with the most widely used method~\cite{van2011rich, JG2014, Seguin2018}, which consists in assigning the weight of a connection by averaging over all weights in the dataset related to that connection, and that typically gives more homogeneous weight distributions and destroys completely the disparity property. Finally, the complementary cumulative strength distribution $P_c (s)$, inset in Fig.~\ref{fig:weightdists}a, shows a homogeneous region for low values of strength and a tail that decays as a power-law with characteristic exponent close to $-2$.

The relation between strength and degree $s (k)$ (Fig.~\ref{fig:weightdists}c) also exhibits a power-law form with average exponents $\mu=0.8 \pm 0.1$ and $\mu=0.71 \pm 0.07$ within subjects in the UL and HCP datasets, respectively. This is in stark contrast with complex networks in other domains, which typically display a superlinear relation. Hence, hubs in brain connectomes accumulate less strength than expected given their connectivity. This effect can be explained by two observations. First, higher degree nodes have a longer average distance to nearest neighbors $\langle d_{\rm nn} \rangle (k)$, as shown in Fig.~\ref{fig:weightdists}e, which increases as a power-law with positive average exponents $\mu=0.36 \pm 0.04$ and $\mu=0.281 \pm 0.019$ in the UL and HCP datasets respectively, indicating that their connections are at all length scales, while lower degree nodes tend to connect preferentially with closer neighbors. Second, the longer the distance between connected nodes, the smaller the weight of the connection between them, as proved by the power-law decay of weight as a function of Euclidean distance $w(d) \approx d^{\mu}$ with $\mu=-1.12 \pm 0.13$ for the UL Dataset, shown in Fig.~\ref{fig:weightdists}d and Fig.~SF9, and $\mu=-2.0  \pm 0.1$ for the HCP Dataset, shown in Fig.~SF63. This inverse relationship can be explained by an economy principle related with longer connections requiring more resources and incurring higher costs for setting up and maintenance than shorter ones.

Finally, the complementary cumulative distance distribution $P_c (d)$ between connected pairs of regions, shown in Fig.~\ref{fig:weightdists}f, is slightly more heterogeneous than the distribution of distances between all nodes in the connectome, although both are limited by the size of the brain as an evident physical constraint.

\subsection{Navigating human connectomes by combining weights and spatial distances}
In the previous section, we proved that the weights of the connections in human brain connectomes and the spatial distances that they cover are coupled. It is then natural to study their interplay in navigation protocols taking into account both magnitudes.

In the standard deterministic greedy routing protocol \cite{kleinberg2000navigation,kleinberg2006complex}, a message navigates the brain to reach a destination from a source by selecting as a transient stop the connected neighbor with the minimum spatial distance to the destination, from where the process is repeated. In this process, each node only requires information about its direct neighboring ROIs, and their distances to the final destination can be easily calculated from the coordinates in the embedding space. This makes greedy routing a distributed strategy much more computationally efficient as compared to routing through topological shortest paths. 

\begin{figure}[t]
    \centering
    \includegraphics[width=0.45\textwidth]{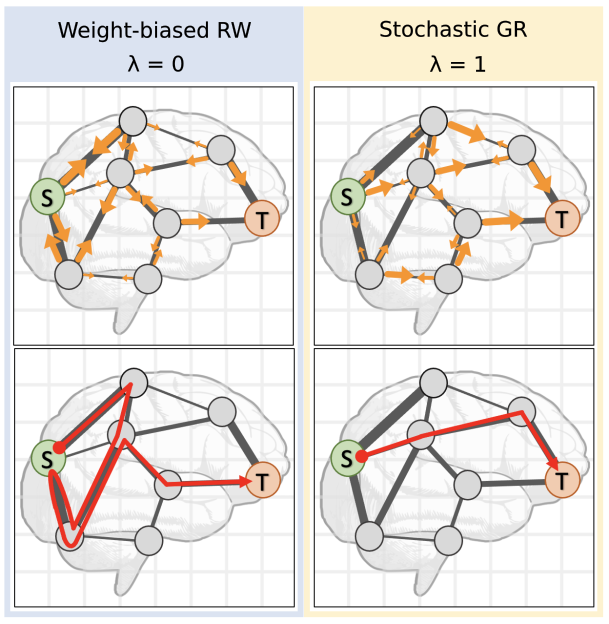}
    \caption{\textbf{Extreme cases of the navigation protocol}. Top row: the probability of transition from one node to its neighbor is represented by an orange arrow. A bigger arrow corresponds to a larger probability. Bottom row: examples of a path taken by the message in each case. The left column corresponds to the case where $\lambda=0$, i.e. weight-biased random walk, and the right column to the case where $\lambda=1$, i.e. stochastic greedy routing. The source node is depicted in green and the target node in orange. }
    \label{fig:nav_scheme}
\end{figure}

To take weights into account, we propose a framework that combines a stochastic greedy routing based on Euclidean distances and a weight-biased random walk, such that messages are preferentially sent along paths with larger connection weights and to nodes closer to the target in the embedding space. This accounts for the expectations that messages are more likely to travel through channels with more nerve fibers and that nearby nodes are connected with higher probability. Altogether, this defines a family of routing models that we term `high weight short distance' (HWSD) routing processes. To decide to which neighbor the information should be transferred to in the next step of a HWSD routing process, we implemented a probability of transition in which distances obtained from weights and spatial distances in the Euclidean embedding of the brain are balanced according to a parameter $\lambda$. Namely, the transition probability from node $i$ to its neighbor $j$ when traveling to target $z$ is
\begin{equation}
    P_{\lambda}(j\ | i,z)=\exp \left[{-(\lambda\cdot d^{\rm{E}}_{jz}+(1-\lambda)\cdot d^w_{ij})}\right]\frac{1}{Z^z_i},
    \label{eq:p2}
\end{equation}
where $Z^z_i=\sum_j \exp\left[{-(\lambda\cdot d^{\rm{E}}_{jz}+(1-\lambda)\cdot d^w_{ij})}\right]$ is a normalization factor, $d^{\rm{E}}$ is the Euclidean distance between the centers of the regions, and $d^{w}$ is the distance obtained by transforming weights such that higher weights correspond to shorter weight distances, see Methods~\footnote{We also tried another framework in which we took as spatial information the total distance from the current node to the target while traveling through neighbor $j$, $d^{\rm E}=d^{\rm{E}}_{ij}+d^{\rm{E}}_{jz}$. The results obtained were qualitatively similar to the ones we got using Eq.~\eqref{eq:p2}, but using only $d^{\rm{E}}_{jz}$ allows for easier comparison with standard greedy routing.}. By gradually changing the value of $\lambda$, the dynamics shifts from a weight-biased random walk at $\lambda=0$, where the movement is guided by the local information provided by the weights of the connections between ROIs, to a stochastic greedy routing at $\lambda=1$, where the dynamics is controlled by spatial distances, i.e. map navigation. See Fig.~\ref{fig:nav_scheme} for an illustrative scheme. 

The proposed procedure is inherently probabilistic and therefore it can take an excessive number of steps to reach the destination. Thus, we applied a time-out (measured in maximum number of steps) after which we assume that the message has died out before reaching the target. We performed the navigation for time-out values of 1000, 2500, 5000, 10000 and 30000.

\begin{figure*}[t]
    \centering
    \includegraphics[width=\textwidth]{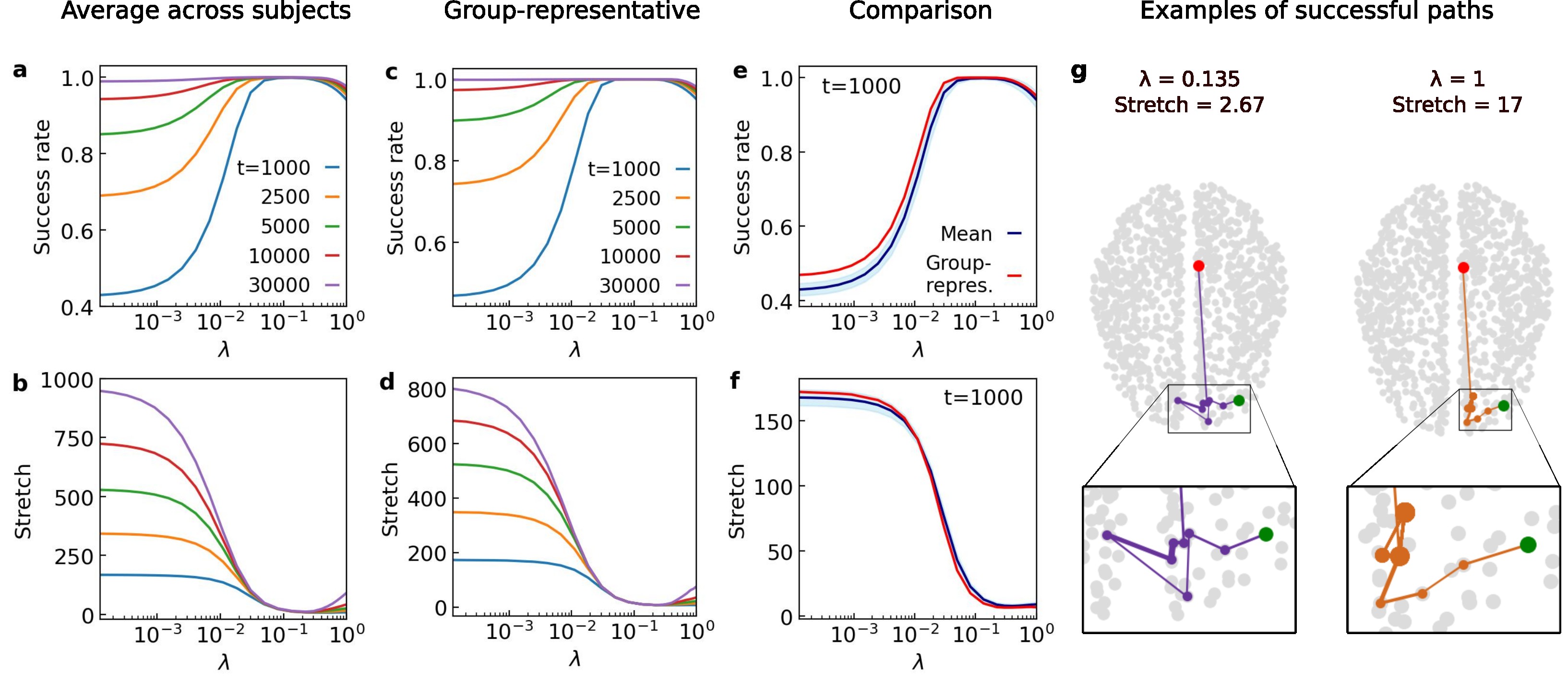}
    \caption{\textbf{a)} Mean success rate and \textbf{b)} mean stretch across the 40 subjects of the UL Dataset. Each curve shows the outcome for a specific value of time-out. \textbf{c)} Success rate and \textbf{d)} stretch for the navigation of the group-representative connectome for different values of time-out. \textbf{e)} and \textbf{f)} for t=1000, comparison of the success rate/stretch for the group-representative connectome (in red) and the mean success rate/stretch of all subjects (in blue). The blue-shadowed area corresponds to two times the standard deviation between subjects. \textbf{g)} Path taken from ROI 154 to ROI 20 of the group-representative connectome for $\lambda=0.135$ --sweet-spot region-- (in purple), and $\lambda=1$ (in orange). The source node is represented in red, the target node in green. In the insets below the size of the nodes is proportional to the number of times the message visits them.}
    \label{fig:nav_all}
\end{figure*}

To evaluate the performance of the protocols, we studied two standard metrics widely employed to assess the efficiency of greedy routing navigation, namely, the success rate and the average stretch. On the one hand, the success rate is the fraction of paths that reach the ROI target successfully, when considering all possible source/target pairs of ROIs in the connectome. In deterministic greedy routing, messages can get stuck in a loop and fail to reach the destination and thus the success rate can be less than one. The stochastic nature of our approach ensures that all messages eventually reach the destination, but not all messages succeed within the time-out we introduced. On the other hand, the stretch of a successful path is calculated as the ratio between the number of links in the path and the number of links in the topological shortest path between source and target. The average stretch is computed over all successful paths. Additionally, we calculated the average transmission cost of successful paths using either Euclidean or weight distances. A paths' transmission cost under a certain routing strategy measures the expected distance that a message has to travel to move along a certain path, and was introduced in \cite{AK2019}, see Methods. Figs.~\ref{fig:nav_all}a and~\ref{fig:nav_all}b provide, respectively, the mean success rate and mean average stretch across the 40 subjects of the UL dataset, while the results for the HCP dataset are reported in Fig.~SF64. The Euclidean and weight distances transmission costs are shown in Figs.~SF10 and SF64. 

Figure~\ref{fig:nav_all} shows that there exists a maximum in the success rate around $\lambda=0.1$ regardless of the time-out value~\footnote{In order to assess to which extent the outcome varies across realizations, we calculated 100 different instances of the navigation protocols in the $\lambda$-spectrum for subject 0 in the UL dataset and time-outs 1000 and 10000. We found that the coefficient of variation between realizations is $c_v \simeq 10^{-3}$. Consequently, and given that we focus on the average outcome of the dataset, running only one realization per subject for each time-out does not introduce any substantial bias to the results.}. In addition, the stretch and the transmission costs have a minimum in the same $\lambda$ region. This means that the combination of both weights and spatial distances to guide the navigation results in maximum navigability, impossible to reach considering only weights or only distances. The same behavior was obtained using the HCP dataset, although the maximum, and specially the minima, are less pronounced, see Fig.~SF64. We report the value of the optimal $\lambda$ for each time-out in Table~\ref{tab:lambdas}. When considered independently, stochastic greedy routing navigation is more efficient than the weight-biased random walk strategy unless the time-out is extremely high, as expected.

The existence of a sweet spot for navigability, reachable by combining weights and distances at approximately $\lambda \simeq 0.1$, may imply that, in order to obtain more successful routes, the weight distances are more relevant than the Euclidean distances, since Euclidean distances are multiplied by a factor $\lambda \simeq 0.1$ whereas weight distances by ${1-\lambda \simeq 0.9}$. However, the mean value of the average weight distances across subjects in the UL dataset is $6.80 \pm  0.12$~mm and the mean average Euclidean distance is $74 \pm 3$~mm, see Figs.~SF25 and SF26. Then, the relative importance between the Euclidean and the weight average distances is ${(74 \cdot 0.1)/(6.8 \cdot 0.9) \simeq 1.21}$, see Fig.~SF27. For the group-representative connectome this ratio is ${(73.8 \cdot 0.1)/(6.83 \cdot 0.9) \simeq 1.2}$. Thus, the optimal performance of the navigation protocol is obtained when spatial distances and weight distances are balanced and approximately equally relevant. For the specific ratios depending on time-out see Table~\ref{tab:lambdas}.

The maximum navigability region is well reproduced in the group-representative connectome, with outcomes that are very similar to the mean of all subjects. It is important to highlight that all navigation magnitudes are accurately described by the group-representative for all time-outs, Figs.~\ref{fig:nav_all}c,d. Figures~\ref{fig:nav_all}e and~\ref{fig:nav_all}f show the comparison of the UL group-representative outcome and the mean results of the dataset for $t=1000$, while Figs.~SF20-SF24 show the results for all time-out values and Figs.~SF72-SF76 display the results for the HCP dataset. We also found that as the time-out increases, the difference between the stretch curves becomes more pronounced and the group-representative starts outperforming individual subjects. The transmission costs behave similarly to the stretch. 

In Fig.~\ref{fig:nav_all}g, we show the example of two successful paths taken to travel from ROI 154 to ROI 20 in the group-representative connectome for $\lambda=0.135$ (close to the optimal $\lambda$) and $\lambda=1$. This is a clear example that illustrates the advantage of incorporating both weights and distances when navigating the network rather than solely considering distances. In the zoomed area shown at the bottom, the ROI's sizes are directly proportional to the number of times the message reaches them. The increased sizes for $\lambda=1$ indicate that, when only considering distances, the message gets stuck for a long time between three of the nodes in the path until it can finally escape.

\subsection{Assessing the role of weights in maximum navigability}
To understand the interplay between weights, topology, and spatial distances in our navigation protocol, we compared the navigation results with those obtained using two null models of the mapped connectomes. The Coordinate-Preserving Weight-Reshuffling (CP-WR) null model preserves the geometry and topology of the network, but reassigns the weights of the connections at random. The Coordinate-Reshuffling Weight-Preserving (CR-WP) null model preserves the topology and weights, but randomizes the geometry by reassigning the position of the nodes in the Euclidean embedding space. In both cases, the inverse relationship between weights and distances is disrupted. 

Since the reshuffling of weights or coordinates in the null models is performed at random, the results can vary depending on the specific realization. In order to investigate whether the choice of the seed affected the outcome, we calculated 50  realizations for subject 34 of the UL dataset and $t=5000$. We found that there is little variation between realizations, with a maximum coefficient of variation $c_v \simeq 7 \times 10^{-2}$, Figs.~SF34 and~SF41. Consequently, and given that we focus on the qualitative differences between null models and original networks and not in the quantitative values, we observed that the results were not altered when running only one realization per subject for each time-out.

\begin{figure*}[t]
    \centering
	\includegraphics[width=\textwidth]{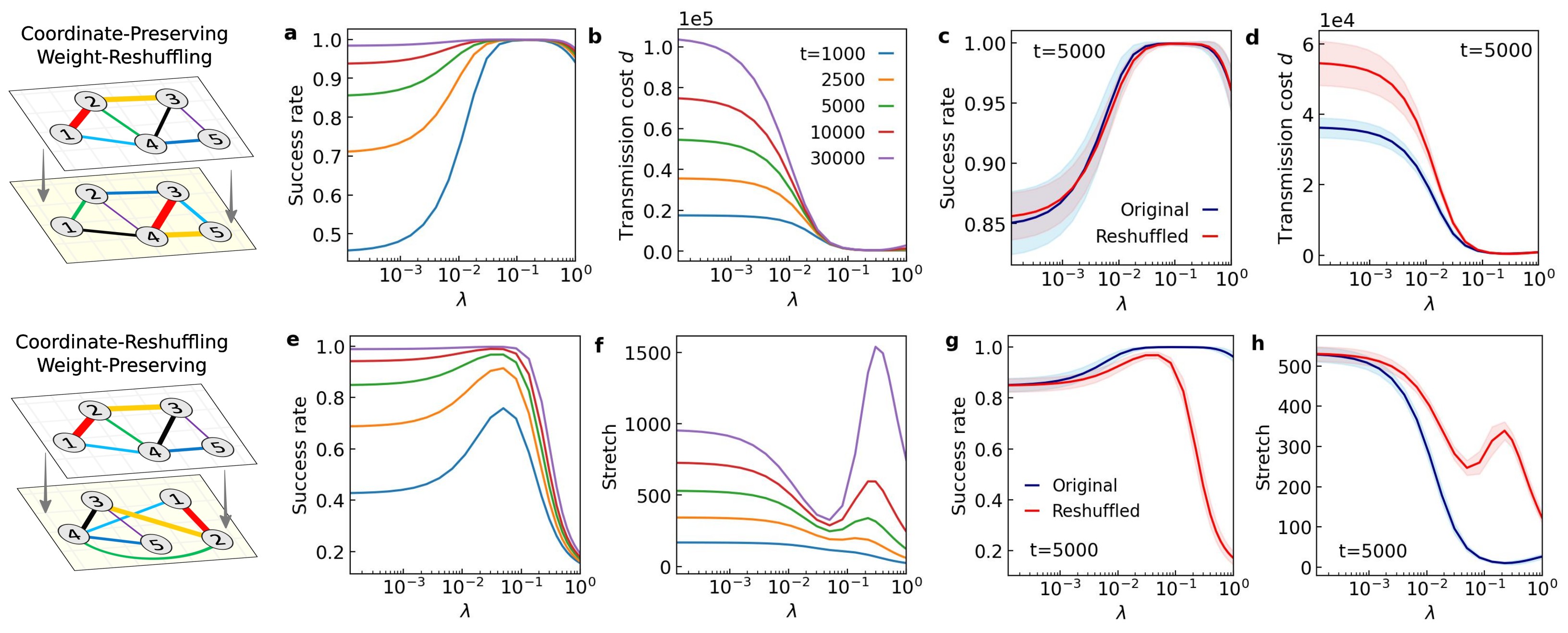}
    \caption{\textbf{Top row: Navigation after weight reshuffling.} \textbf{a)} Success rate and \textbf{b)} Euclidean distance transmission cost for different values of time-out. \textbf{c)} and \textbf{d)} for t=5000, comparison of the mean success rate/Euclidean distance transmission cost between the original connectomes (in blue) and the reshuffled networks (in red). The shadowed areas correspond to two times the standard deviation between subjects. \textbf{Bottom row: Navigation after position reshuffling.} \textbf{e)} Success rate and \textbf{f)} stretch for different values of time-out. \textbf{g)} and \textbf{h)} for t=5000, comparison of the mean success rate/stretch between the original connectomes (in blue) and the reshuffled networks (in red). The shadowed areas correspond to two times the standard deviation between subjects.}
    \label{fig:resh}
\end{figure*}

\textit{Coordinate-Preserving Weight-Reshuffling null model.} The success rate, the average stretch, and the transmission cost of weight distances are maintained for all values of $\lambda$, as shown in Figs.~\ref{fig:resh}a,c and SF31. By contrast, the difference in the transmission cost of Euclidean distances between the original connectomes and the randomized surrogates increases as $\lambda$ decreases, Fig.~\ref{fig:resh}b,d, as a consequence of the disruption of the weight distance correlation. In the original connectome, when the navigation protocol prioritizes traveling through connections with high weights, neighbors that are close in space are selected. The reshuffling of weights disrupts this correlation and, consequently, the messages traverse a longer total distance in the reshuffled network than in the original one. Interestingly, in the optimal region of the $\lambda$ spectrum all analyzed magnitudes behave similarly in the CP-WR null model and in the original network.  Finally, even though here we show the results for t=5000, the pattern of the outcomes is similar for the various time-outs studied in this work, Figs.~SF29-SF33. Therefore, the relative difference between the curves for different time-outs remains unaltered after the reshuffling, even for the Euclidean distance transmission cost, see Figs.~\ref{fig:resh}a,b and SF10,~SF28. 

We obtained similar results when, instead of reshuffling the weights of the connections, we assigned the weight-distances at random from a uniform distribution around the mean, see Fig.~SF48. In addition, we checked that when weights are constant, we obtain the results for a stochastic greedy routing scaled by $\lambda$.

\textit{Coordinate-Reshuffling Weight-Preserving null model.}
In the reshuffled surrogates, the use of only Euclidean distances to guide navigation procured worse results than the weight-biased random walk, see Fig.~\ref{fig:resh}e. Precisely, the amount of successful paths for $\lambda=1$ was below $20\%$. Another interesting finding is that, even in coordinate-reshuffled connectomes, there is a range of $\lambda$ values (close to the sweet spot in the original connectome) for which the success rate has a maximum, although it does not reach the level of the sweet spot. Therefore, it is clear that complementing the weight-biased random walk with information about spatial distances, even after repositioning nodes, improves navigation results. To understand this improvement one has to keep in mind that the introduction of any spatial distances to calculate the probability of transition increases the chances of traveling through small-weight connections. This diversifies the choice of neighbors as compared to the bare weight-biased random walk and increases the chances of reaching more targets within the time-out. At larger values of $\lambda$, the stochastic greedy routing based on spatial distances dominates, and the fact that distances are not coupled to the topology causes the success rate to drop steadily. The results of comparing the average stretch and the transmission costs before and after the reshuffling of positions, Figs.~\ref{fig:resh}h and SF10,~SF35, show that the CR-WP randomization leads to a performance decline in all magnitudes.

In light of these results, we can state that the key ingredients that maximize navigability in the original connectomes are the coupling between topology and geometry as well as the presence of weights, but not their specific values. The weights are important to randomize the choice of neighbors as compared to the stochastic greedy routing based solely on spatial distances, such that new routes can be found. This `weight facilitation' reinforces the maximum navigability phenomenon in human brain connectomes as akin to a type of stochastic resonance effect where weights play the constructive role of noise.

\subsection{Maximum navigability is resilient under targeted attack}

\begin{figure*}[t]
    \centering
    \includegraphics[width=0.95\textwidth]{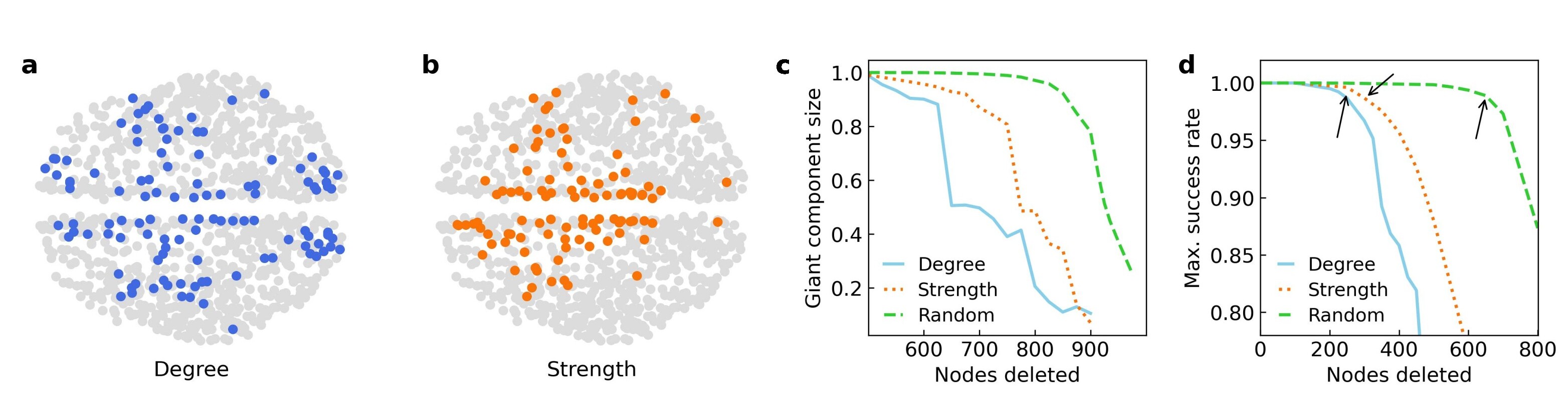}
    \includegraphics[width=0.95\textwidth]{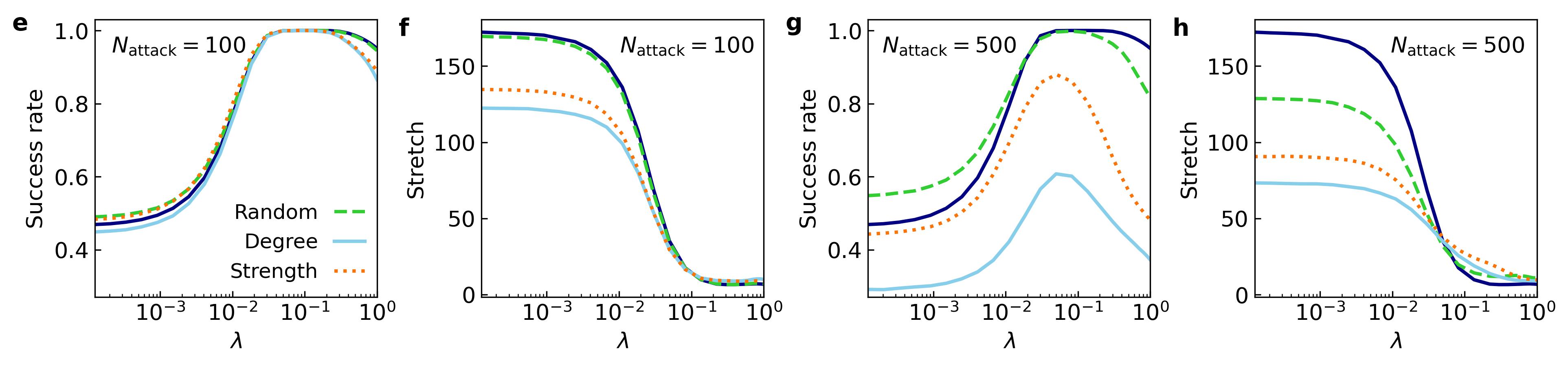}
    \caption{\textbf{Effects of targeted attacks on the UL group-representative connectome.} \textbf{a)} ROIs deleted in the targeted attack on nodes with highest degree and \textbf{b)} with highest strength for $N_{\rm attack}=100$.  \textbf{c)} Evolution of the relative size of the giant connected component of the network as nodes are deleted. \textbf{d)} Evolution of the maximum success rate reached as a function of the nodes deleted. The arrows point at each of the points where the success rate does not reach anymore the value it had in the sweet spot for the original network. \textbf{e, g)} Success rate and \textbf{f, h)} average stretch after deleting 100 and 500 nodes of the network, respectively. The dark-blue curve represents the original results, the dashed green line the results after deleting nodes at random, the light blue plain curve corresponds to the targeted attack on nodes with highest degree and the dotted orange curve to the attack on nodes with highest strength.}
    \label{fig:attack}
\end{figure*}

Finally, we evaluated the resilience of the group-representative connectome of the UL Dataset under perturbations affecting the number of accessible ROIs. More specifically, we analyzed its navigability for $t=1000$ after inflicting two types of ROI-deletion targeted attacks, namely on ROIs with the highest degrees (HD), Fig.~\ref{fig:attack}a, or on ROIs with the highest strength (HS), Fig.~\ref{fig:attack}b. As a baseline reference, we also computed the effects of attacking randomly selected nodes and averaged the results over 10 realizations (RD). In addition, the validity as an archetype of the group-representative was corroborated by the analysis of seven individuals in the UL dataset. 

As expected, the differences between the navigation of the original network and of the perturbed one increase as the number of attacked regions grows, regardless of the type of attack. Also, the targeted attacks are not able to disconnect the remaining ROIs, which typically form a robust giant connected component. For instance, it is necessary to remove approximately 60\% of the ROIs in decreasing order of degree to break off the connectome. In the rest of the cases, this percentage rises to 70\% in descending order of strength, and to 90\% when nodes are deleted at random, see Fig.~\ref{fig:attack}c. 

Figures~\ref{fig:attack}e-h display the results for $N_{\rm attack}=100,500$. Even after removing $10\%$ of the ROIs, the success rate maintains the original performance and the existence of the maximum navigability sweet spot, and the three types of damage produce very similar results to the original network. In fact, the random attack produces results very similar to the unperturbed network even after the deletion of 500 nodes. However, the HD and HS perturbations affect quicker the navigability of the connectomes and after the removal of 500 nodes all magnitudes are severely disrupted. Specifically, slight deviations can be observed in the success rate for HS, Fig.~SF43a, with results getting worse in the stochastic greedy routing limit and slightly better in the weight-biased random walk extreme since the attractors of the random walk are removed and, consequently, the probability of the message getting stuck in some nodes decreases. Regarding HD, the success rate decreases for most values of $\lambda$ as we progressively delete nodes. The average stretch has two different behaviors on each side of the minimum both for HS and HD: to the right it does not change approximately until $250$ nodes are removed and to the left it improves as we delete nodes, Figs.~SF42b,~SF43b. For the HS and HD, the Euclidean distances transmission cost has a behavior similar to the stretch, and the weight distances transmission cost worsens for all values of $\lambda$ as the number of deleted nodes increases, as seen in Figs.~SF42,~SF43. 

Interestingly, none of the attacks affected significantly the efficiency of the HWSD protocol in the $\lambda$ region that defines the maximum navigability sweet spot in the original group-representative connectome. Even after the removal of the top $25\%$ of nodes with highest degree, $30\%$ of nodes with highest strength and $65\%$ at random, there is a region in the spectrum of $\lambda$ where navigation is still optimal, see Fig.~\ref{fig:attack}d and Figs.~SF42-SF44. In other words, by combining weights and distances to guide the navigation, a message can be delivered to almost every ROI remaining in the connectome even after serious damage.

\section{Discussion and conclusions}

The brain, and the nervous system in general, is plastic and has the ability to adapt to external or internal changes by reorganizing its structure and functions. Apart from extreme events, for instance injuries such as stroke, the reshaping of neural connections and their strength can also be induced by processes such as memory storage and learning. Not all the reconfiguration processes have the same cost or occur at the same time scale. Additionally, changing the hard wiring of the brain, that is, the large scale connectivity structure of the connectome, is very difficult and can only happen to a very limited extent, while changing the soft wiring, understood as a reconfiguration of weights such that new knowledge or experiences are fixed, is an everyday process.

The results of the analyzes reported in this work show that weights are necessary to achieve maximum navigability in brains, defining a sweet spot in which human connectomes become maximally navigable and achieve full communication efficiency in a very robust fashion. The results hold on human brain connectomes at the personalized level and are well captured in group-representatives, which serve as suitable archetypes for analyzing the navigation and resilience of the connectomes within the cohort. We specifically designed the group-representative connectomes to maintain not only the topological but also the weighted properties of the network. 

Interestingly, the particular configuration of link weights is not important but rather they avail in randomizing the choice of neighbors as compared to the stochastic greedy routing based solely on spatial distances, such that new routes can be found. This result is congruent with the hypothesis that efficient communication between brain regions should not depend crucially on memory or learning processes that can dynamically change the configuration of weights at short time scales. Moreover, the fact that the maximum navigability phenomenon in human brain connectomes is independent of the particular configuration of weights, supports an interpretation of the phenomenon as akin to a type of stochastic resonance effect where weights play the constructive role of noise. The combination of information about weights and distances seems, thus, a particularly convenient way to route information in the brain. This not only happens in terms of efficiency but also of resilience. Indeed, targeted attacks on the brain regions with high degree or high strength were not able to disconnect the remaining ROIs and maximum navigability was achieved even under serious damage. 

Our work provides a theoretical framework for the investigation of how the geometry of brain connectomes and their weighted structure, the \textit{hard} and \textit{soft} wiring of the brain, are tailored to the needs of information transmission between connected brain regions. The broad spectrum of stochastic routing protocols introduced here may serve to guide future experimental investigations of these mechanisms. For now, our results illustrate that information about both weights and spatial distances are required to achieve maximally efficient and robust communication, needed to support ultrafast responses of the brain to external and internal stimuli, and ultimately to behavior and cognition.

Yet many specific questions remain unanswered. For instance, about the evolutionary processes that control the specific spatial positioning of brain regions, and how specific configurations of weights in connectome links are related with different functional responses in the communication sweet spot. Regarding greedy routing protocols, future work will need to address thoroughly the biological plausibility of these models to explore large-scale communication in the brain. Although there is some evidence that targeted information processing may play a role in brain communication dynamics \cite{ciocchi2015science}, an empirical justification that regions or neurons could possess knowledge on the spatial positioning of their neighbors in relation to a target region or neuron is still lacking (see Ref. \cite{Seguin2018} for a detailed discussion).

\section{Materials and Methods} \label{Methods}

\subsection{Data description}
We used the human connectomes of a total of 84 healthy adult subjects in two different datasets. The two datasets were previously used in~\cite{Zheng2020} to assess the multiscale organization of the connectivity structure of human brain connectomes. The UL dataset comprises connectomes of 40 healthy subjects (16 females) around 25 years old obtained from diffusion spectrum MRI images. Neural fibers connecting pairs of regions were tracked by following directions of maximum diffusion. The data was processed using The Connectome Mapper Toolkit \cite{tourbier2020connectomicslab}. The HCP Dataset~\cite{van2012human}, used to cross-validate the results, consists of 44 connectomes (31 females) aged 22-35 years old obtained from T1-weighted and corrected diffusion-weighted magnetic resonance images. All the connectomes were reconstructed by using deterministic streamline tractography methods, and the parcellation of the cortex was defined according to the Desikan-Killiany atlas.  A full description of the data acquisition and processing can be found in~\cite{Zheng2020}. We found similar results for both cohorts.

All connectomes encompass both hemispheres and comprise typically $N=1014$ nodes without the brainstem region. There is a negligible variability in the number of nodes across subjects due to the removal of regions that were isolated in the original dataset or that became isolated after the removal of the brainstem, and of regions that were only connected to themselves by a self-loop. 

In this work, we used the information about the fiber density as connection weight. The weight $w_{ij}$ of the connection between nodes $i$ and $j$ is calculated as
\begin{equation}
	w_{ij}=\frac{2}{S_i+S_j} \sum_{f \in F_e} \frac{1}{l(f)}, 
\end{equation}
where $S_i$ and $S_j$ are the areas of ROIs $i$ and $j$ respectively, $F_e$ is the set of all fibers connecting ROIs $i$ and $j$, and $l(f)$ is the length of the fiber  between ROIs $i$ and $j$. The connection weight between a pair of brain regions is thus defined as the number of streamlines connecting them per unit surface, corrected by the length of the streamlines. The aim of these normalizations is to control for the variability in cortical region size and the linear bias toward longer streamlines introduced by tractography algorithms.

\subsection{Disparity of weights} 
The local heterogeneity of link weights for each node can be assessed with the function \cite{barthelemy2003spatial}
\begin{equation}
	\Upsilon_i (k) \equiv kY_i(k)= k \sum_j p_{ij}^2= k \sum_j (w_{ij}/s_i)^2
\end{equation}
where $w_{ij}$ is the weight of the link and $s_i$ the strength of the node. The function $Y_i(k)$ has been extensively used in various research fields and it is known in complex networks as the disparity measure. When the strength of the node is distributed homogeneously between its connections $\Upsilon_i (k)=1$. On the other hand, when just one link hoards all the strength of the node $\Upsilon_i (k)=k$. The null model mentioned in the caption of Fig.~\ref{fig:weightdists} is given by $\Upsilon(k)<\mu(\Upsilon_{\rm null}(k))+2 \sigma(\Upsilon_{\rm null}(k))$ \cite{MAS2009}, where 
\begin{equation}
	\mu(\Upsilon_{\rm null}(k))=\frac{2k}{k+1},
\end{equation}
and
\begin{equation}
	\sigma^2(\Upsilon_{\rm null}(k))=k^2\left(\frac{20+4k}{(k+1)(k+2)(k+3)}-\frac{4}{(k+1)^2}\right).
\end{equation}

\subsection{Weight distances} 
When performing the navigation we want to combine Euclidean distances and weights in one equation. Therefore, they must be the same kind of magnitude and we have to transform weights into weight distances. The fiber density is a proximity measure --proximity measures are often interpreted as the information flow or traffic capacity that can travel through a connection. Thus, to define weight distances from weights, it is necessary to use a proximity-to-distance mapping. This mapping is designed to transform big weights into small weight distances and small weights into big weight distances. We used the proximity-to-distance mapping from \cite{AK2019}, 
\begin{equation}
	d_{ij}=A\ln(1/w_{ij}), 
\end{equation}
where $d_{ij}$ is the weight distance, $w_{ij}$ is the weight of the connection and $A=1$ mm is a parameter introduced to ensure correct dimensionality. This mapping was also used in other previous studies \cite{JG2014,AK2016}.

\subsection{Lambda values} Since we want to explore the whole navigation spectrum, i.e. from only considering weights to only considering Euclidean distances, we will take $\lambda$ values going from 0 to 1 with logarithmic spacing from $e^{-9}$ to $e^{-1.5}$ and regular spacing from $0.3$ to $1$. 

\subsection{Transmission cost} The immediate transmission cost of a node $i$ is the expected distance that a message has to travel to move to a neighbor of $i$ when going to the target $z$ \cite{AK2019}. The Euclidean distance immediate transmission cost and weight distance immediate transmission cost are computed respectively with
\begin{equation}
    c^{\rm{trans}\ E}_{\lambda}(i,z)=\sum_j P_{\lambda}(j\ | i,z)\cdot d^{\rm{E}}_{ij}
    \label{eq:ctransd}
\end{equation}
and
\begin{equation}
    c^{\rm{trans}\ w}_{\lambda}(i,z)=\sum_j P_{\lambda}(j\ | i,z)\cdot d^{w}_{ij},
    \label{eq:ctransw}
\end{equation}
where $j$ are the neighbors of node $i$, $P_{\lambda}(j\ | i,z)$ is the probability of transition from node $i$ to its neighbor $j$ when going to node $z$ for a certain value of $\lambda$, and $d_{ij}$ is the distance from node $i$ to its neighbor $j$.

The total transmission cost of a path is computed by adding the immediate transmission cost of all nodes visited through the path. Likewise, the total transmission cost in a network is calculated as the average of the transmission cost for all node pairs.

\subsection{Sweet spot region magnitudes}
In Table~\ref{tab:lambdas} we show the $\lambda$ values in the center of the sweet spot region averaged over all subjects in the UL Dataset for each time-out. We also display the ratio between Euclidean and weight distances for these $\lambda$. 

\vspace{0.75cm}
\begin{table}[h!]
    \centering
\begin{tabular}{c|c|c}
      time-out & $\lambda$ & $[\langle d_E \rangle \cdot \lambda]/[\langle d_w \rangle \cdot (1-\lambda)]$ \\ \hline 
1000  & $0.129 \pm 0.018$ & 1.61\\ 
2500  & $0.12 \pm 0.04$ & 1.48\\ 
5000  & $0.10 \pm 0.04$ & 1.21 \\ 
10000  & $0.08 \pm 0.04$ & 0.94\\ 
30000  & $0.07 \pm 0.04$ &  0.82 \\ 
\end{tabular}
    \caption{Mean $\lambda$ values for the maximum in the success rate and its standard deviation across subjects in the UL Dataset for the five time-outs studied. The third column shows the scaled ratios between the mean average Euclidean distance ${\langle d_E \rangle=74}$ mm and the mean average weight distance ${\langle d_w \rangle=6.8}$ mm. }
    \label{tab:lambdas}
\end{table}

\section*{Data and code availability}
The weighted human connectomes will be available upon publication via Zenodo. Our source codes will be available upon publication via GitHub. 

\section*{Declaration of competing interests}
The authors declare no competing interests.

\section*{Acknowledgments}
We thank Patric Hagmann and Yasser Alem\'an G\'omez for sharing the data, and Mari\'an Bogu\~n\'a for useful comments. L.B. acknowledges support from an FPU21/03183 grant from the Spanish Government. J.S. was supported by funded by MCIN/AEI/10.13039/501100011033 under projects PID2019-108842GB-C21 and PID2022-137713NB-C22, and by the Generalitat de Catalunya under project 2021-SGR-00450. M.A.S. acknowledges support from grants PID2019-106290GB-C22 and PID2022-137505NB-C22 funded by MCIN/AEI/10.13039/501100011033, and from grant number 2021SGR00856 by the Generalitat de Catalunya.

\bibliography{reference}
	
\newpage\hbox{}\thispagestyle{empty}\newpage



\section*{Supplementary material}

# Optimal navigability of weighted human brain connectomes in physical space

## Supplementary Material

Laia Barjuan,[1,2] Jordi Soriano,[1,2] and M. Ángeles Serrano[1,2,3,*]

[1]*Departament de Física de la Matèria Condensada,*
*Universitat de Barcelona, Martí i Franquès 1, E-08028 Barcelona, Spain*
[2]*Universitat de Barcelona Institute of Complex Systems (UBICS), Universitat de Barcelona, Barcelona, Spain*
[3]*ICREA, Passeig Lluís Companys 23, E-08010 Barcelona, Spain*

## CONTENTS

--------

* marian.serrano@ub.edu

# I. RESULTS FOR UL DATASET

## A. Network properties

| Human | Nodes | Edges | $\langle k \rangle$ | $k_{\max}$ | $\langle s \rangle$ | $s_{\max}$ | $\langle d_{nn} \rangle$ | $\langle d_{nn}^{\max} \rangle$ | $\langle \lambda \rangle$ | Diameter |
|---|---|---|---|---|---|---|---|---|---|---|
| 0 | 1009 | 14470 | 28.682 | 253 | 0.231 | 1.586 | 36.507 | 78.307 | 2.838 | 6 |
| 1 | 1010 | 14406 | 28.527 | 311 | 0.196 | 1.388 | 37.501 | 80.963 | 2.732 | 6 |
| 2 | 1014 | 13671 | 26.964 | 244 | 0.202 | 1.416 | 35.251 | 76.002 | 2.939 | 7 |
| 3 | 1011 | 12991 | 25.699 | 277 | 0.203 | 1.677 | 33.089 | 69.293 | 2.897 | 7 |
| 4 | 1014 | 15879 | 31.320 | 324 | 0.214 | 1.367 | 39.766 | 83.277 | 2.712 | 6 |
| 5 | 1014 | 14340 | 28.284 | 323 | 0.198 | 1.358 | 37.336 | 79.459 | 2.750 | 6 |
| 6 | 1013 | 12660 | 24.995 | 274 | 0.193 | 1.662 | 31.392 | 66.028 | 2.846 | 6 |
| 7 | 1014 | 13474 | 26.576 | 320 | 0.180 | 1.838 | 35.743 | 75.007 | 2.801 | 6 |
| 8 | 1002 | 13910 | 27.764 | 243 | 0.194 | 1.575 | 36.282 | 73.946 | 2.827 | 6 |
| 9 | 1010 | 13496 | 26.725 | 272 | 0.184 | 1.339 | 34.297 | 73.021 | 2.828 | 6 |
| 10 | 1014 | 15222 | 30.024 | 294 | 0.214 | 1.493 | 37.517 | 81.667 | 2.763 | 5 |
| 11 | 1013 | 14695 | 29.013 | 287 | 0.215 | 1.473 | 36.993 | 78.241 | 2.741 | 6 |
| 12 | 1001 | 13933 | 27.838 | 256 | 0.226 | 1.250 | 38.990 | 81.805 | 2.914 | 7 |
| 13 | 1013 | 14409 | 28.448 | 299 | 0.207 | 1.832 | 35.525 | 76.016 | 2.770 | 6 |
| 14 | 1013 | 14959 | 29.534 | 258 | 0.203 | 1.267 | 37.058 | 79.019 | 2.752 | 6 |
| 15 | 1007 | 16208 | 32.191 | 334 | 0.179 | 1.272 | 40.996 | 88.262 | 2.731 | 5 |
| 16 | 1014 | 14539 | 28.677 | 328 | 0.187 | 1.983 | 37.666 | 80.333 | 2.757 | 6 |
| 17 | 1014 | 13901 | 27.418 | 287 | 0.202 | 1.105 | 33.093 | 71.965 | 2.796 | 6 |
| 18 | 1014 | 12418 | 24.493 | 293 | 0.198 | 1.574 | 31.971 | 70.447 | 2.851 | 6 |
| 19 | 1011 | 14883 | 29.442 | 275 | 0.190 | 1.271 | 38.261 | 80.939 | 2.751 | 6 |
| 20 | 1011 | 13201 | 26.115 | 277 | 0.204 | 1.904 | 33.242 | 72.015 | 2.851 | 6 |
| 21 | 1012 | 12004 | 23.723 | 269 | 0.220 | 1.922 | 32.751 | 70.064 | 2.952 | 7 |
| 22 | 1014 | 13519 | 26.665 | 248 | 0.178 | 1.191 | 34.254 | 73.564 | 2.820 | 6 |
| 23 | 1014 | 14709 | 29.012 | 256 | 0.227 | 1.635 | 36.617 | 79.446 | 2.787 | 6 |
| 24 | 1014 | 13661 | 26.945 | 265 | 0.187 | 1.242 | 34.009 | 73.554 | 2.824 | 6 |
| 25 | 1013 | 14802 | 29.224 | 294 | 0.233 | 1.688 | 35.992 | 77.251 | 2.733 | 5 |
| 26 | 1013 | 12942 | 25.552 | 274 | 0.213 | 1.856 | 32.330 | 68.722 | 2.836 | 7 |
| 27 | 1014 | 12483 | 24.621 | 287 | 0.195 | 1.618 | 33.245 | 70.181 | 2.838 | 6 |
| 28 | 1012 | 14812 | 29.273 | 312 | 0.194 | 1.109 | 35.170 | 75.165 | 2.708 | 5 |
| 29 | 1012 | 13601 | 26.879 | 285 | 0.183 | 1.259 | 35.566 | 76.319 | 2.853 | 7 |
| 30 | 1013 | 14177 | 27.990 | 307 | 0.216 | 1.528 | 35.480 | 78.136 | 2.777 | 6 |
| 31 | 1014 | 15641 | 30.850 | 284 | 0.214 | 1.735 | 38.648 | 80.802 | 2.728 | 7 |
| 32 | 1013 | 12875 | 25.420 | 291 | 0.192 | 1.389 | 32.513 | 68.954 | 2.847 | 6 |
| 33 | 1014 | 14510 | 28.619 | 269 | 0.222 | 2.153 | 36.457 | 81.054 | 2.796 | 6 |
| 34 | 1012 | 14344 | 28.348 | 273 | 0.193 | 1.610 | 35.678 | 77.531 | 2.779 | 6 |
| 35 | 1009 | 12460 | 24.698 | 270 | 0.183 | 1.484 | 32.548 | 70.207 | 2.875 | 7 |
| 36 | 1014 | 13978 | 27.570 | 252 | 0.185 | 1.657 | 35.341 | 76.665 | 2.801 | 6 |
| 37 | 1014 | 14282 | 28.170 | 282 | 0.232 | 1.766 | 35.538 | 74.774 | 2.745 | 6 |
| 38 | 1011 | 14001 | 27.697 | 260 | 0.212 | 1.319 | 37.182 | 79.424 | 2.823 | 6 |
| 39 | 1013 | 12599 | 24.875 | 293 | 0.173 | 2.036 | 35.104 | 74.253 | 2.823 | 6 |
| Average | 1012.025 | 13976.625 | 27.621 | 282.5 | 0.202 | 1.546 | 35.573 | 76.052 | 2.805 | 6.1 |
| Maximum | 1014 | 16208 | 32.191 | 334 | 0.233 | 2.153 | 40.996 | 88.262 | 2.952 | 7 |
| Minimum | 1001 | 12004 | 23.723 | 243 | 0.173 | 1.105 | 31.392 | 66.028 | 2.708 | 5 |
| Group-representative | 1014 | 14418 | 28.438 | 372 | 0.267 | 1.639 | 29.5 | 69.694 | 2.703 | 5 |

**ST 1**: Number of nodes, number of edges, average degree, maximum degree, average strength, maximum strength, mean average distance to nearest neighbors, mean maximum distance to nearest neighbors, average shortest path and diameter of the network for all connectomes in the UL Dataset. The last four rows correspond to the mean between all subjects, the maximum of each of the columns, the minimum and the results for the group-representative connectome.

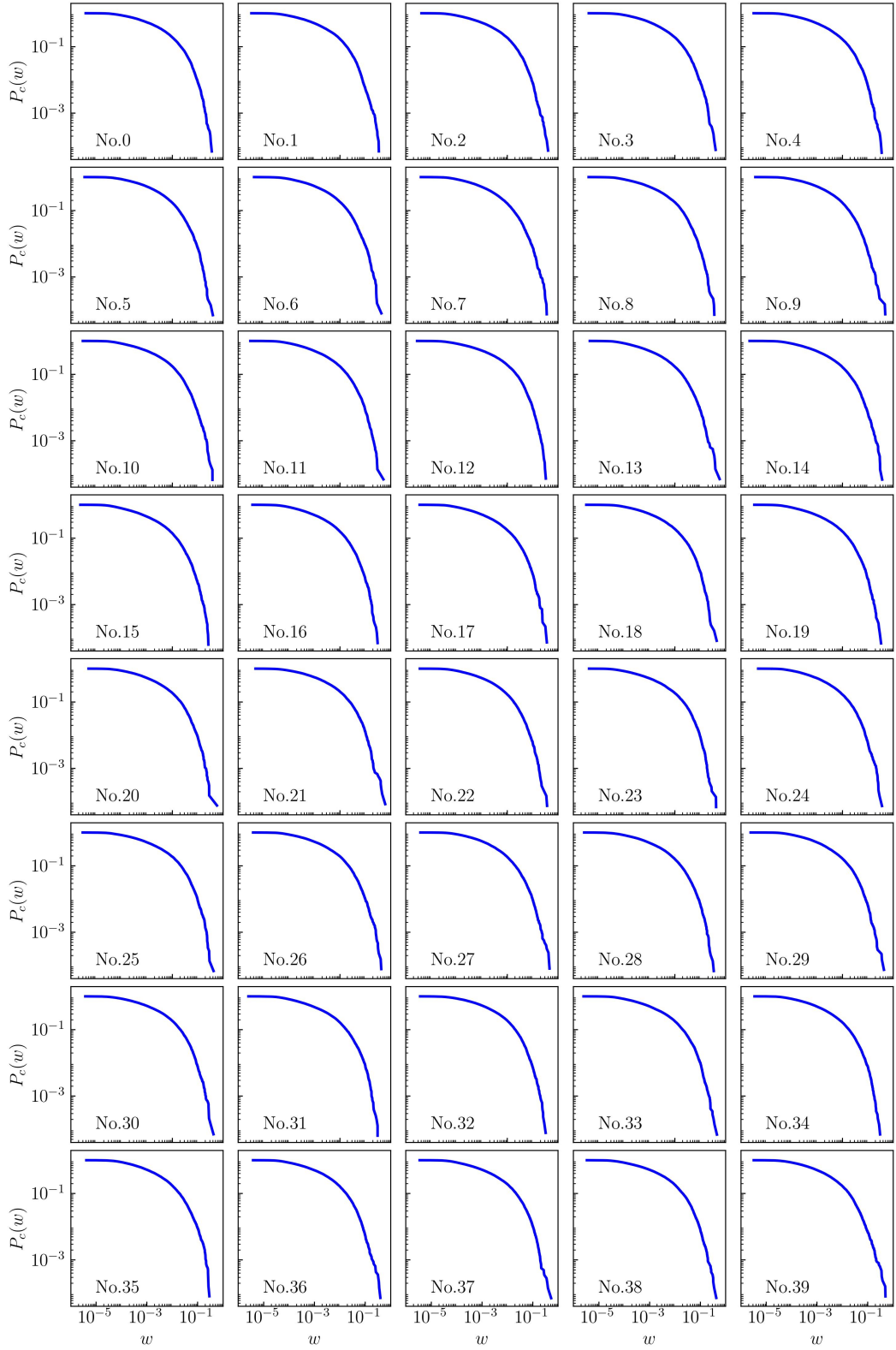

**SF 1**: Complementary cumulative weight distribution $P_c(w)$. Each of the plots corresponds to one subject.

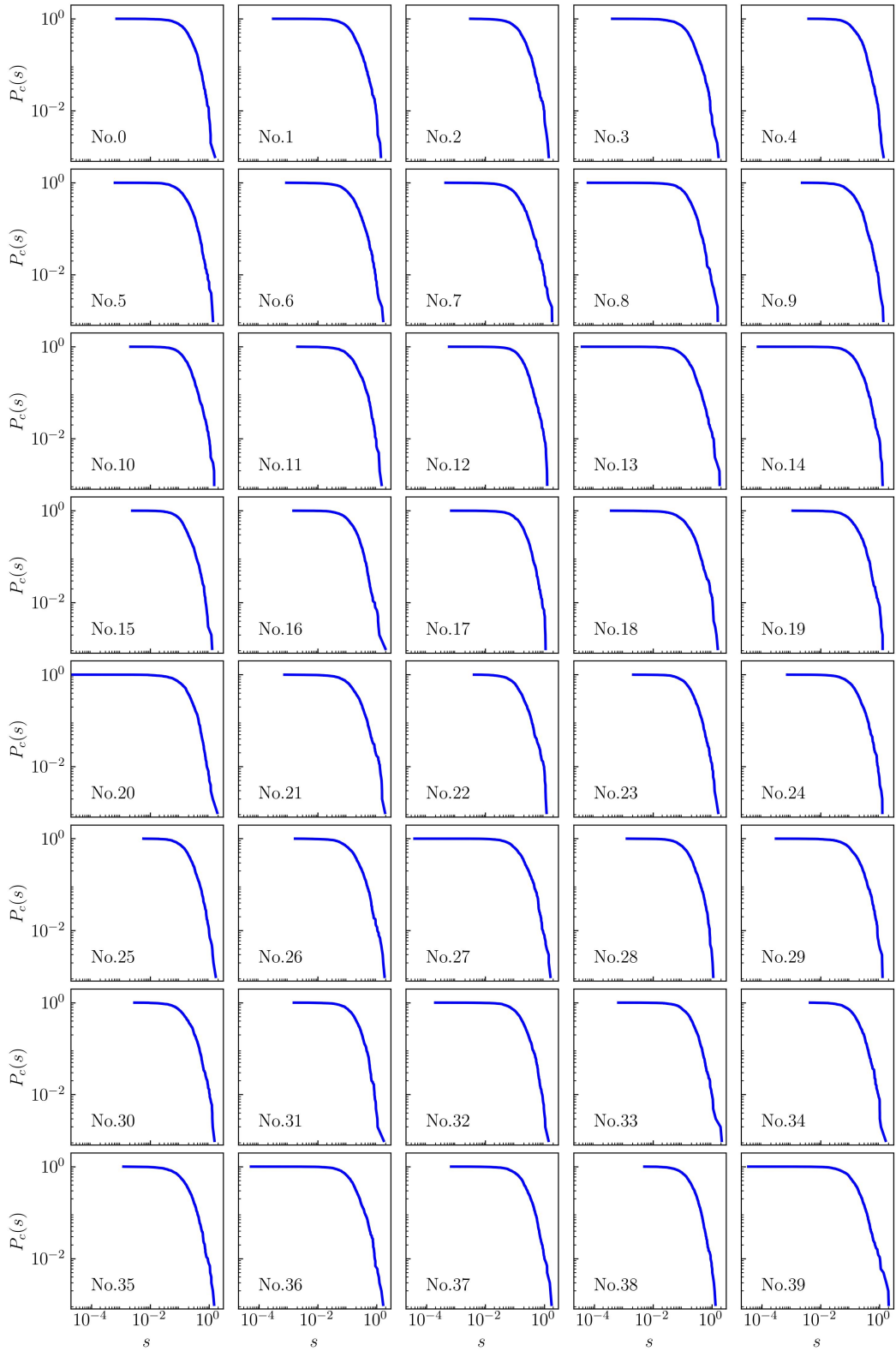

**SF 2**: Complementary cumulative strength distribution $P_c(s)$.

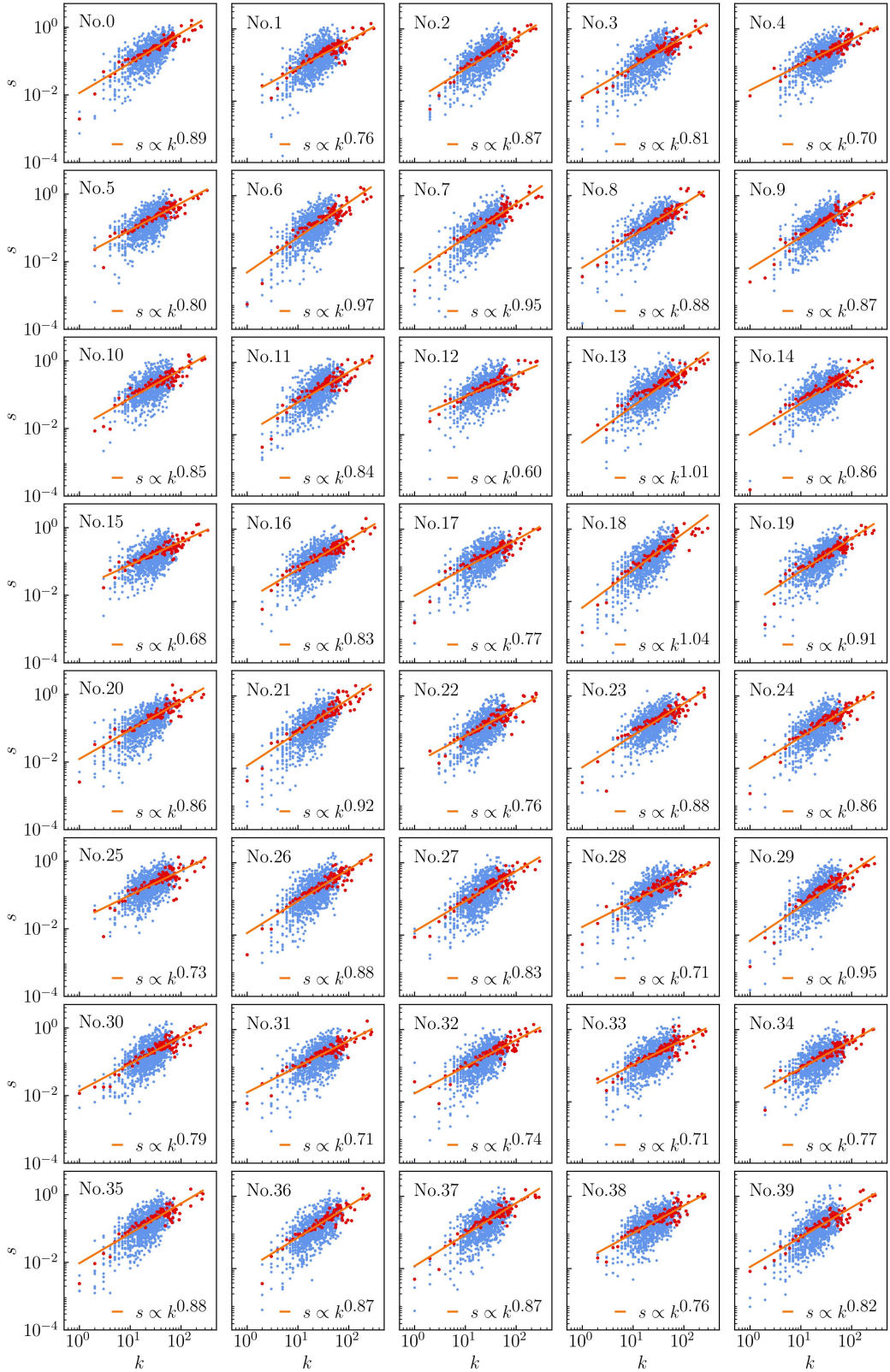

**SF 3**: Degree-strength correlation. Each of the blue dots corresponds to one node of the connectome. The red dots are the average values of strength for each degree. The orange curve is the fit of the averages. Average exponent: $\mu = 0.8 \pm 0.1$.

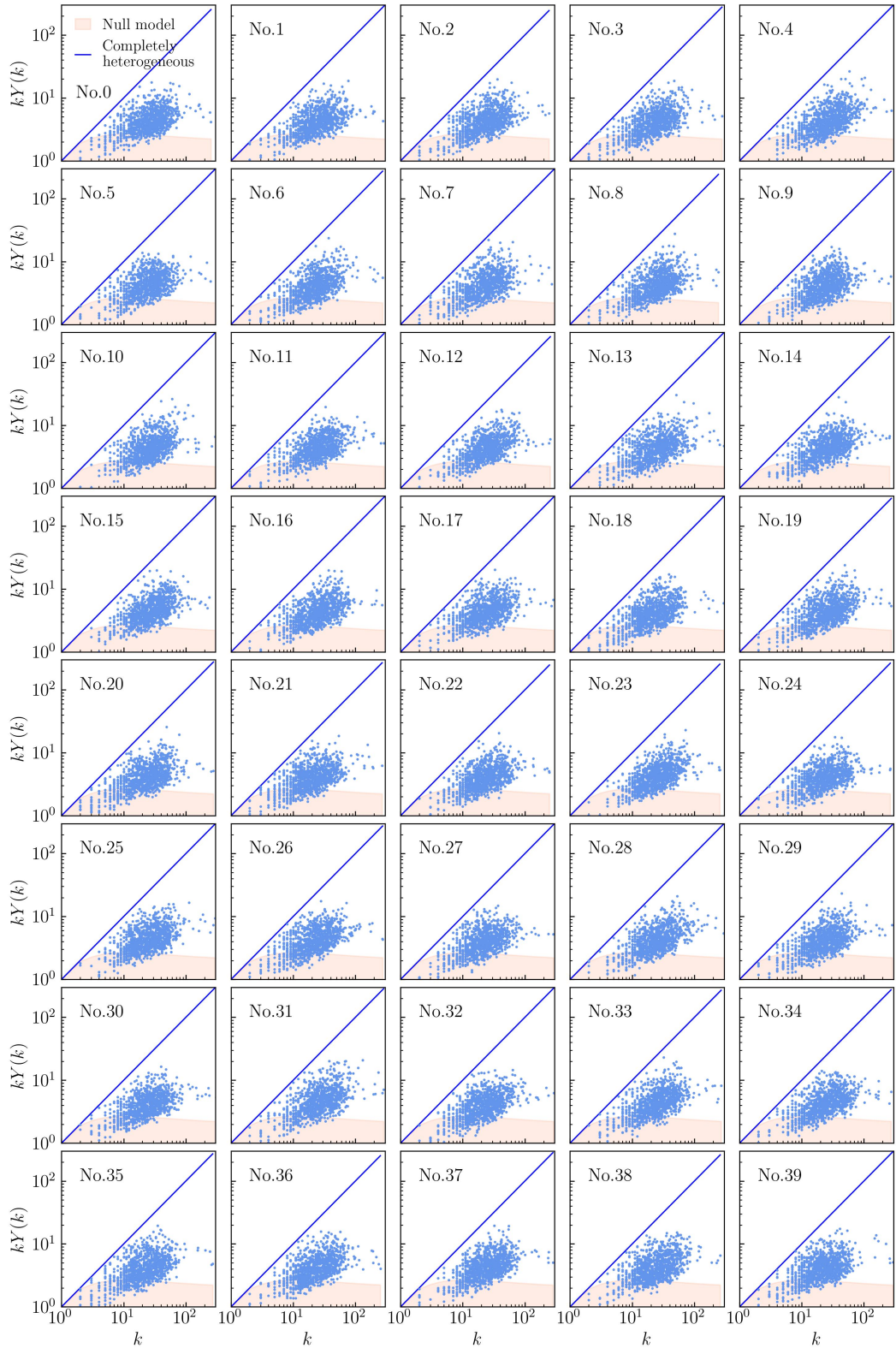

**SF 4**: Disparity measure. The orange area corresponds to the average $+2\sigma$ given by the null model explained in Methods. The blue line represents the completely heterogeneous scenario. Each of the blue dots corresponds to one node of the connectome.

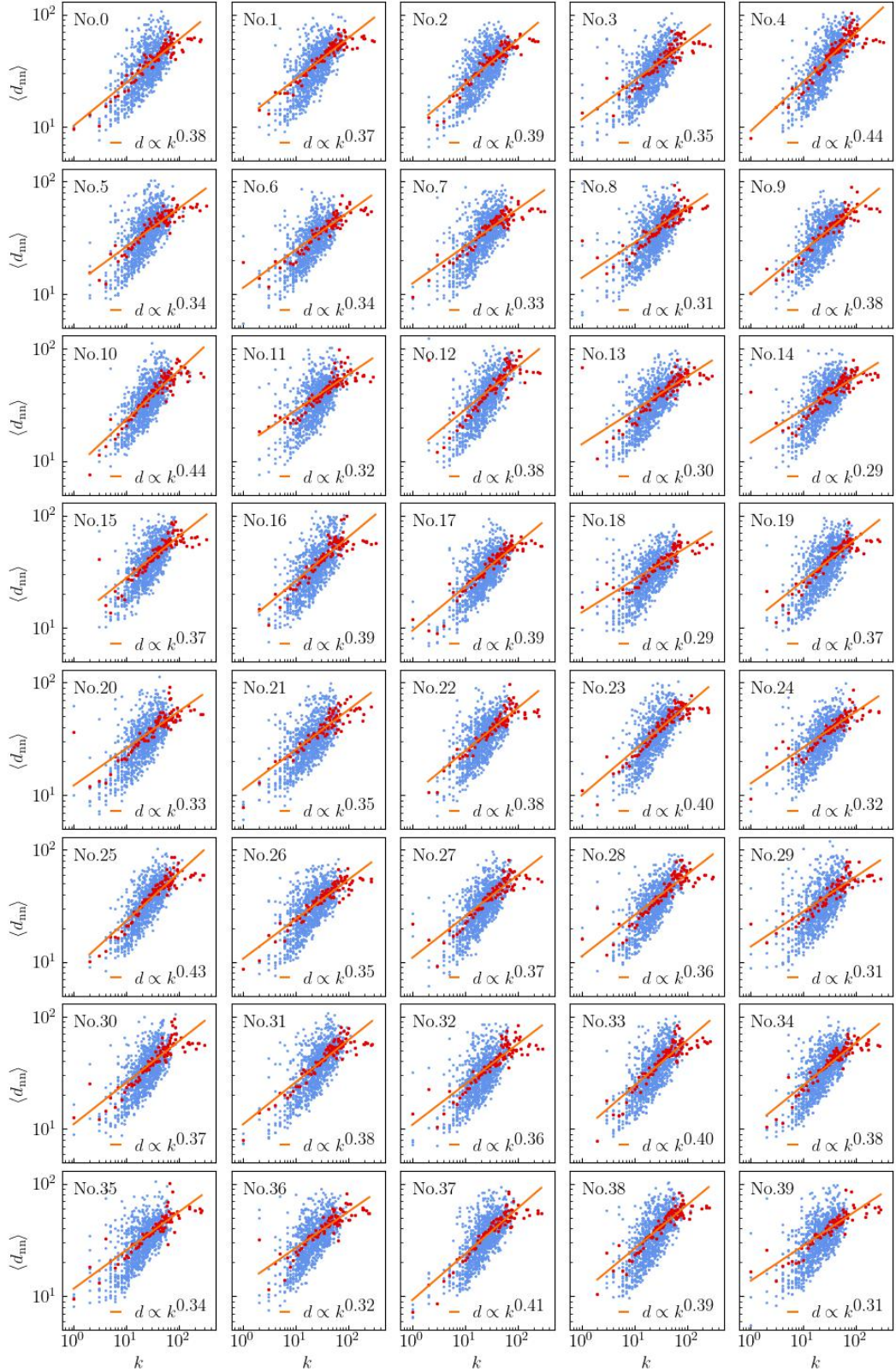

**SF 5**: Average Euclidean distance to nearest neighbours with respect to degree. Each of the blue dots corresponds to one node of the connectome. The red dots are the mean values of Euclidean distance for each degree. The orange curve is the fit of the averages. Average exponent: $\mu = 0.36 \pm 0.04$.

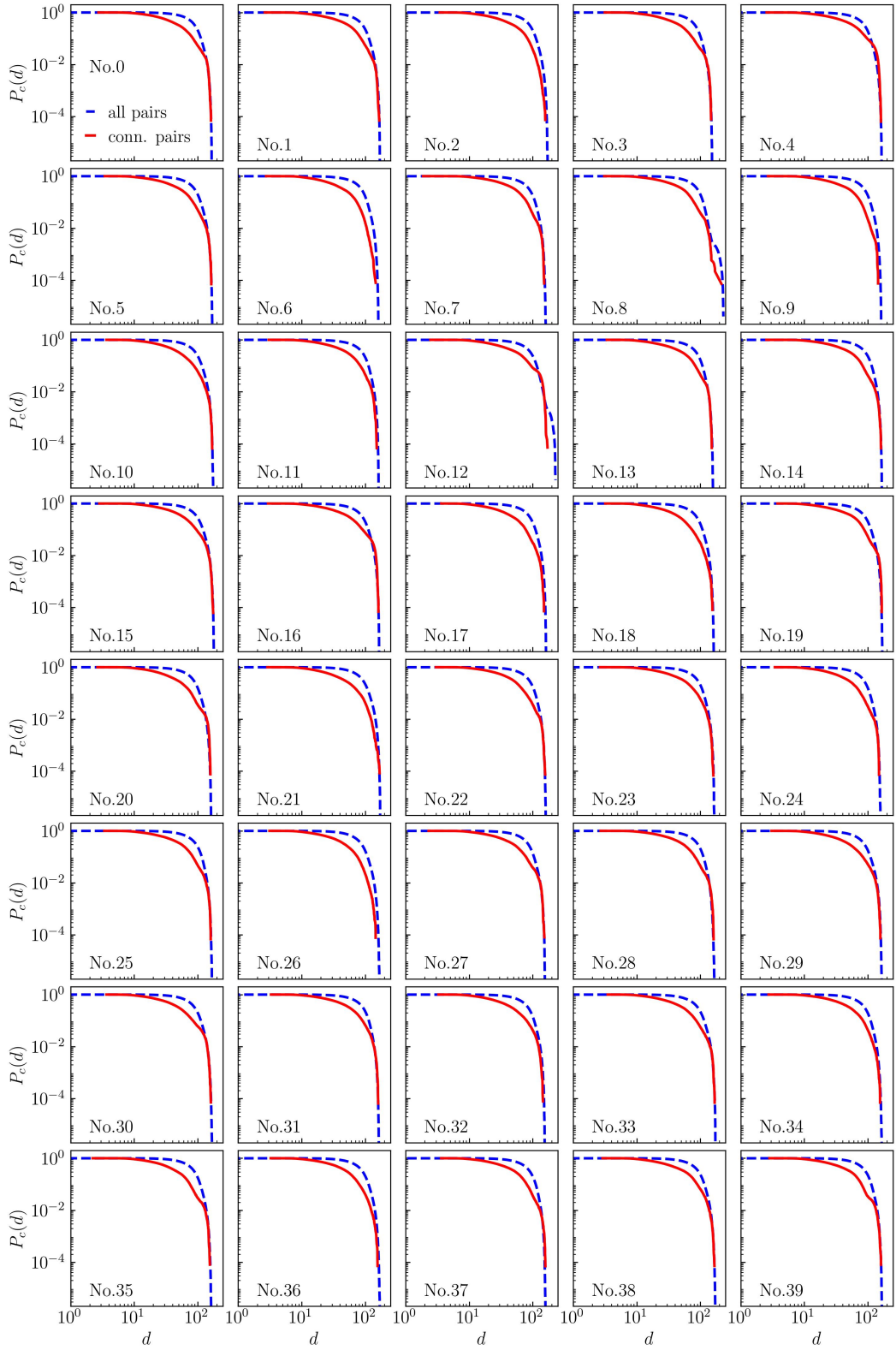

**SF 6**: Complementary cumulative Euclidean distance distribution, $P_c(d)$. The dashed blue curve corresponds to the distribution of distances between all pairs in the network and the red curve to the distances between connected pairs.

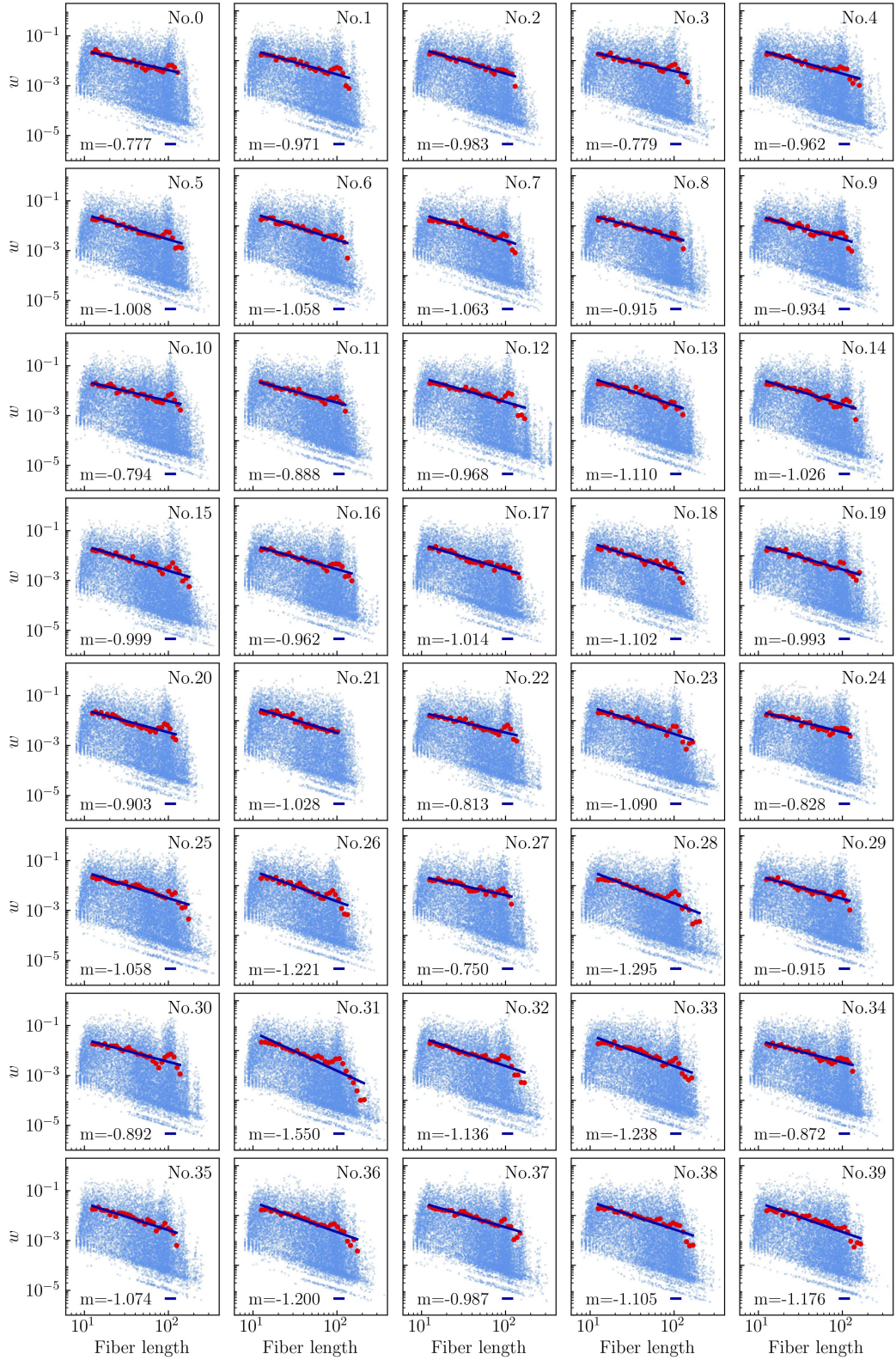

**SF 7**: Relatioship between weight and fiber length. The red dots correspond to the average weight for each of the 30 logarithmically-spaced fiber length bins ranging from flength$_{min}$×1.5 to flength$_{max}$×0.5. The fit of the averaged values is depicted in blue and it has the form $\ln(w) = m\ln(\text{flength}) + n$, with $\langle m \rangle = -1.01 \pm 0.16$.

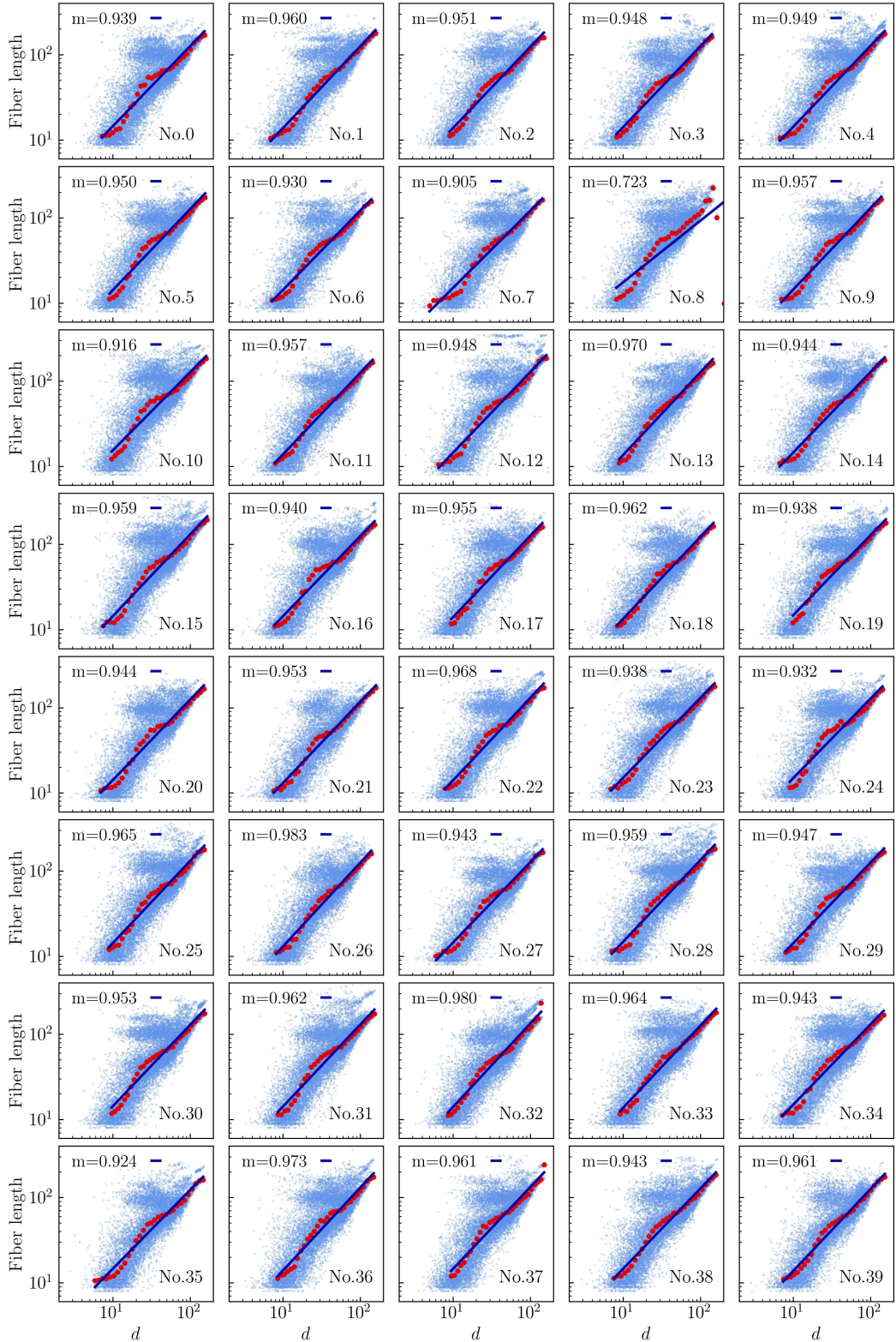

**SF 8**: Relatioship between fiber length and Euclidean distance. The red dots correspond to the average fiber length in each of the 30 logarithmically-spaced Euclidean distance bins ranging from $d_{\min} \times 1.5$ to $d_{\max} \times 0.5$. The fit of the averaged values is depicted in blue and it has the form $\ln(\text{flength}) = m \ln(d) + n$, with $\langle m \rangle = 0.95 \pm 0.04$.

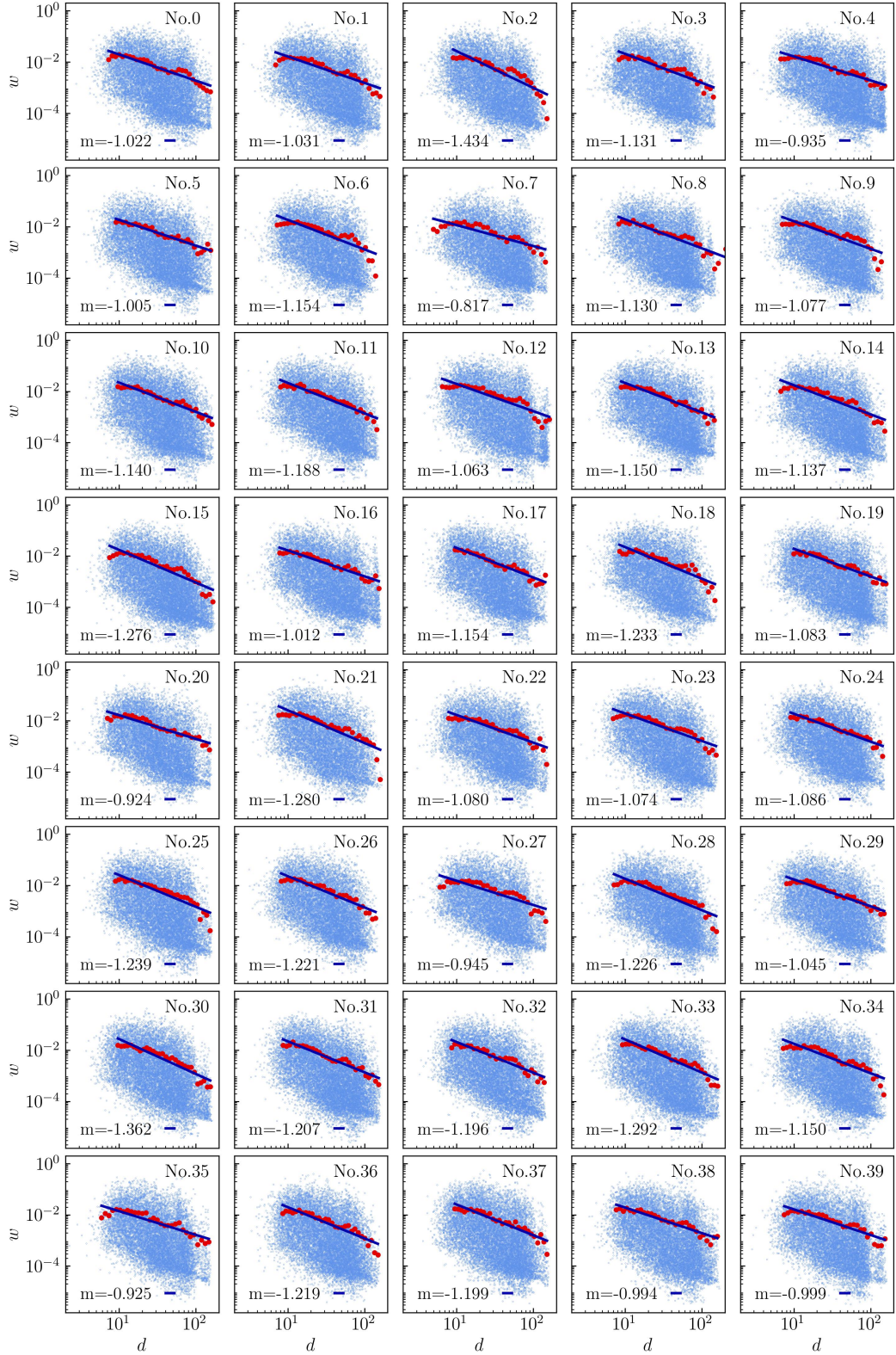

**SF 9**: Relatioship between weight and Euclidean distance. The red dots correspond to the average weight for each of the 30 logarithmically-spaced Euclidean distance bins ranging from $d_{\min} \times 2.5$ to $d_{\max}$. The fit of the averaged values is depicted in blue and it has the form $\ln(w) = m\ln(d) + n$, with $\langle m \rangle = -1.12 \pm 0.13$.

B.   **Navigation results**

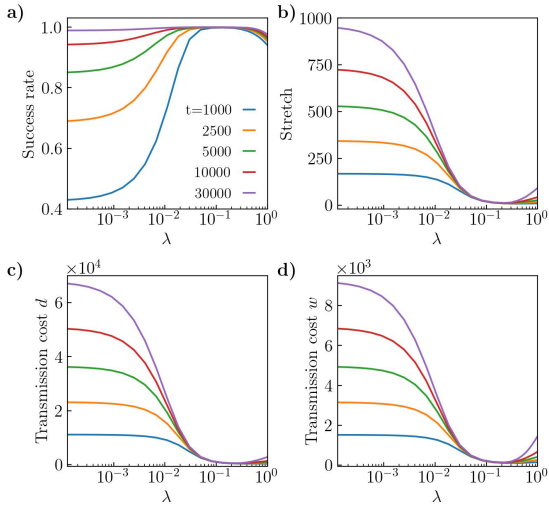

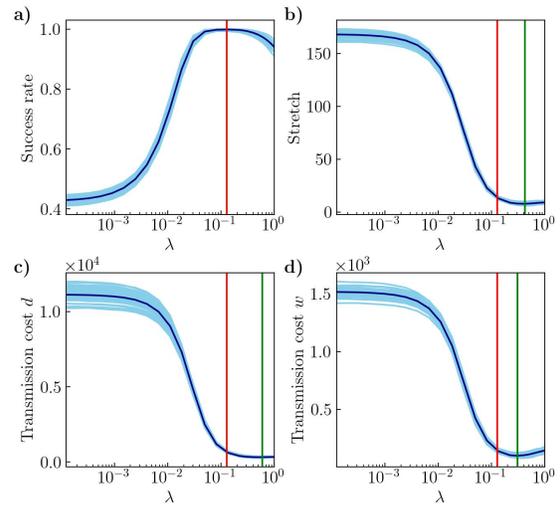

**SF 10**: Mean results across the 40 subjects of the UL Dataset of **a)** success rate, **b)** stretch, **c)** Euclidean distance transmission cost and **d)** weight distance transmission cost. Each curve shows the outcome for a specific value of time-out.

**SF 11**: t=1000. Sky-blue curves show the results for each subject. The dark-blue curve represents the mean. The $\lambda$ for the maximum value of success rate **(a)** is indicated using a vertical red line in all plots, and in the plots of the stretch, Euclidean and weight distance transmission costs **(b,c,d)** the $\lambda$ for the minimum value of the respective magnitude is shown using a vertical green line.

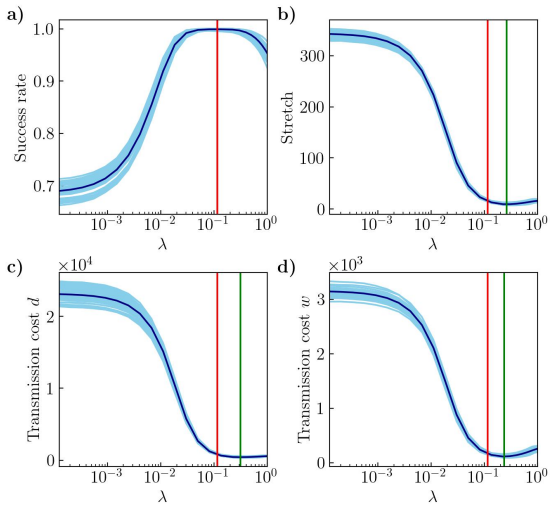

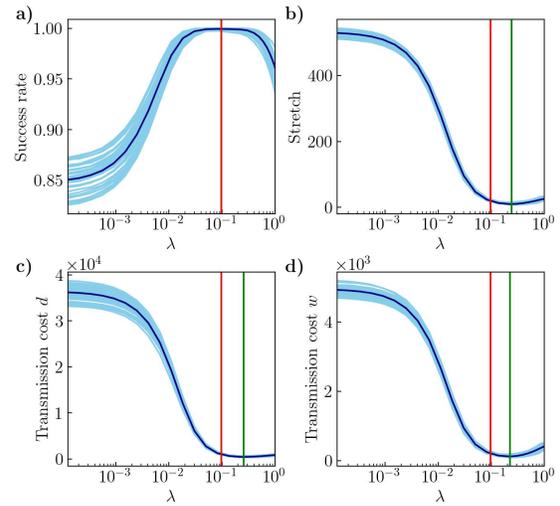

**SF 12**: Same as Fig. **SF 11** but for t=2500.

**SF 13**: Same as Fig. **SF 11** but for t=5000.

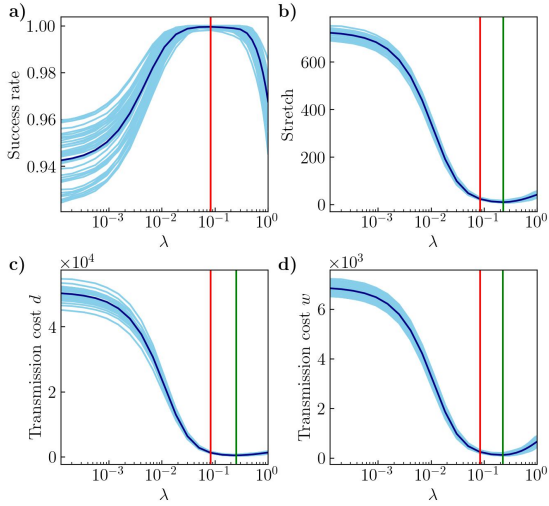
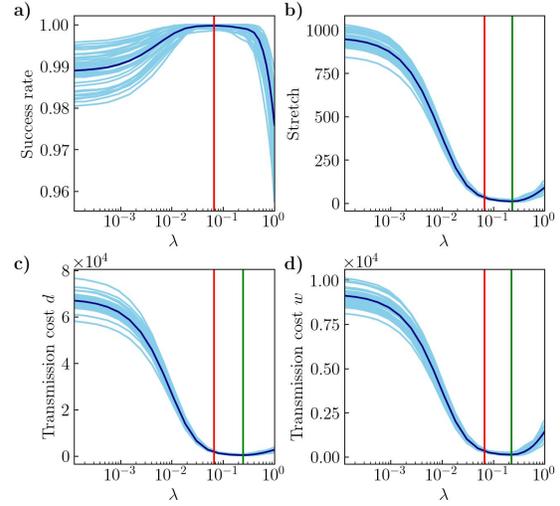

**SF 14**: Same as Fig. **SF 11** but for t=10000.

**SF 15**: Same as Fig. **SF 11** but for t=30000.

## C.  Errors in navigation

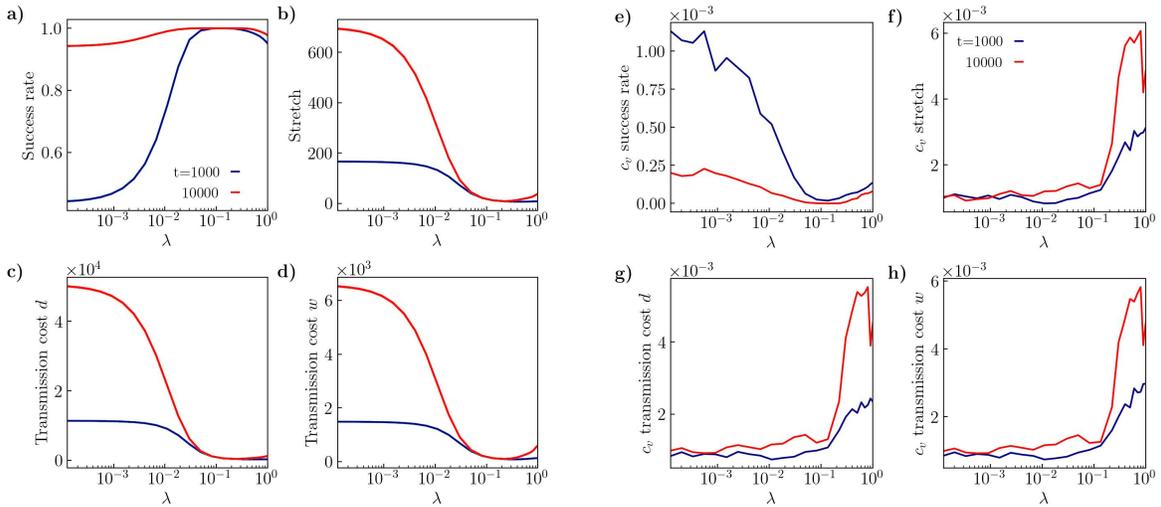

**SF 16**: **a-d)** 100 navigation realizations for subject Human 0 and time-outs 1000 and 10000. The mean of all realizations are represented in blue and red and each of the realizations in light blue and light red for time outs 1000 and 10000 respectively. **e-h)** Coefficient of variation $c_v = \frac{\sigma}{\mu}$, where $\mu$ is the mean of all realizations and $\sigma$ is their standard deviation.

### D. Group-representative connectome properties

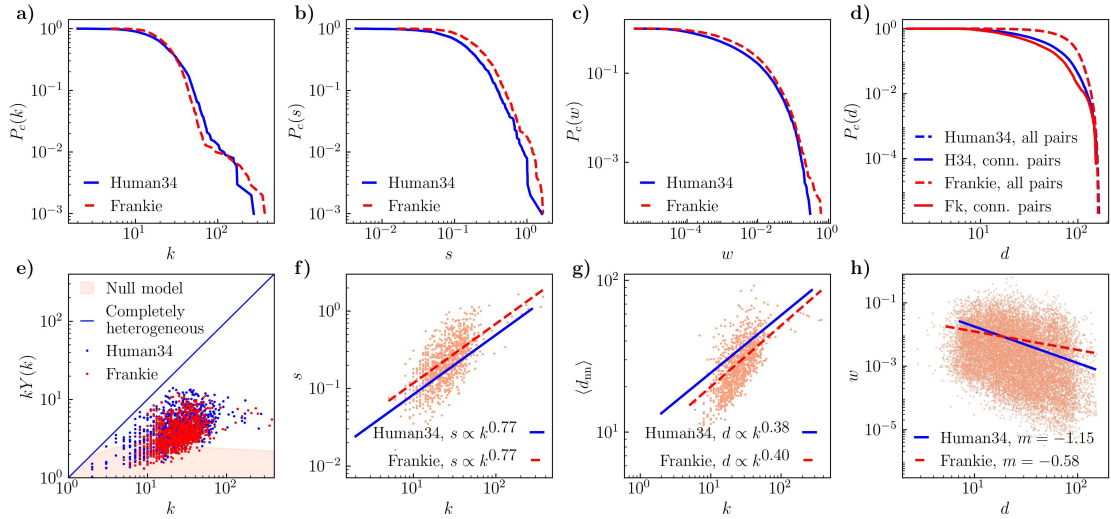

**SF 17**: Comparison of network properties between the group-representative connectome, referred as Frankie, and Human 34. **a)** Complementary cumulative degree distribution $P_c(k)$. **b)** Complementary cumulative strength distribution $P_c(s)$. **c)** Complementary cumulative weight distribution $P_c(w)$. **d)** Complementary cumulative Euclidean distance distribution $P_c(d)$. **e)** Disparity of weights. **f)** Degree-strength correlation. **g)** Average Euclidean distance to nearest neighbours vs degree. **h)** Relatioship between weight and Euclidean distance. In all plots the blue curves show the results for Human 34 and the red ones for the group-representative connectome. The orange dots in **f)**, **g)** and **h)** represent the results for each of the nodes of the group-representative connectome.

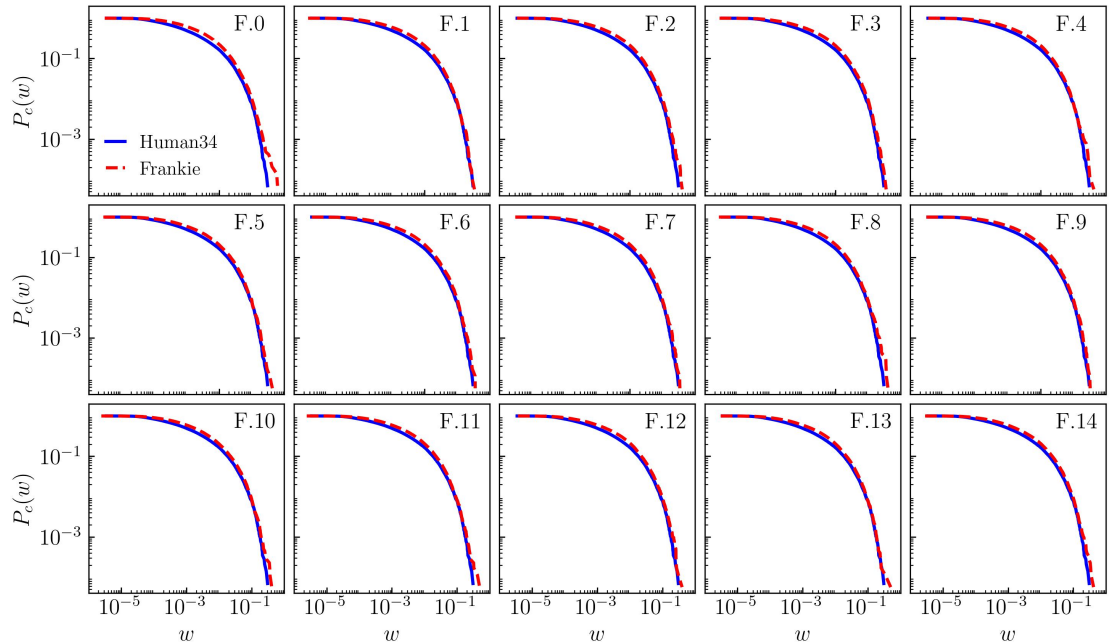

**SF 18**: Complementary cumulative weight distribution for 15 different realizations of the group-representative connectome compared to Human 34. Each of the plots corresponds to one realization of the group-representative.

**E.  Group-representative navigation**

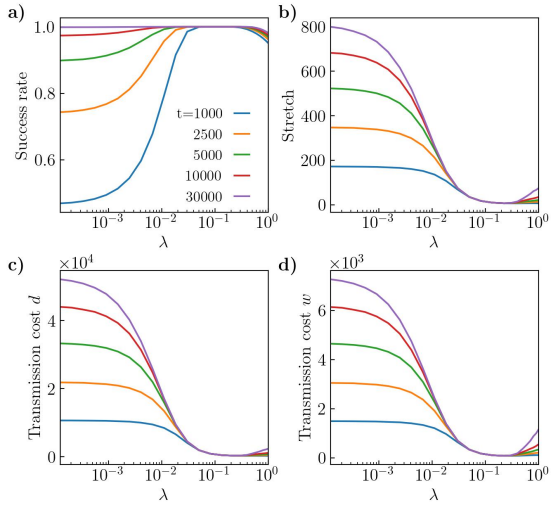

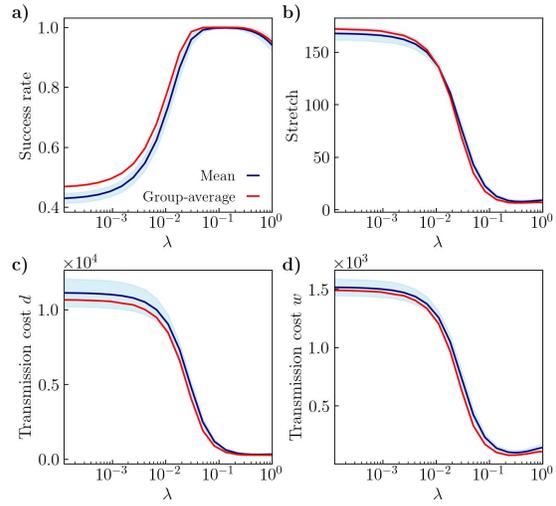

**SF 19**: **a)** Success rate, **b)** stretch, **c)** Euclidean distance transmission cost and **d)** weight distance transmission cost for the group-representative connectome of the UL Dataset. Each curve shows the outcome for a specific value of time-out.

**SF 20**: t=1000. Comparison of the results for the group-representative (in red), here referred as group-average, and the mean results for all subjects (in blue). The blue-shadowed area corresponds to two times the standard deviation between subjects.

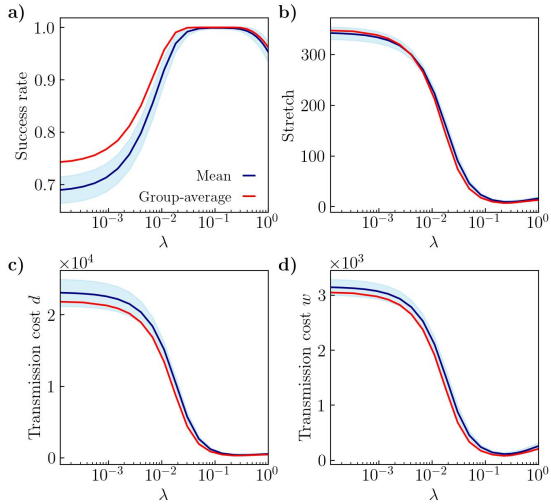

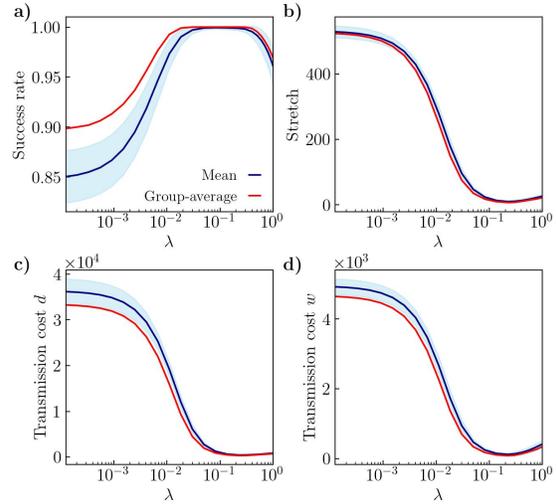

**SF 21**: Same as Fig. **SF 20** but for t=2500.

**SF 22**: Same as Fig. **SF 20** but for t=5000.

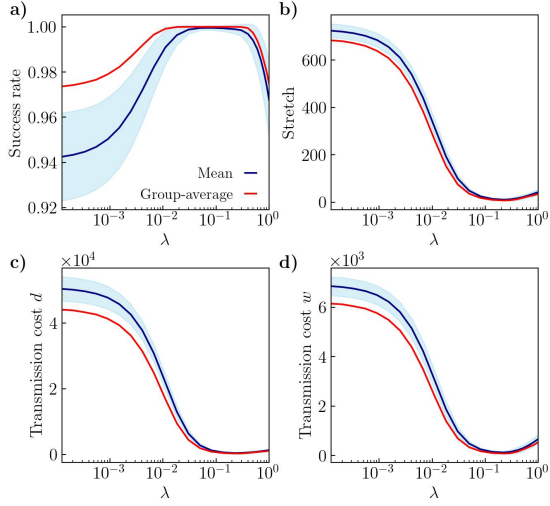 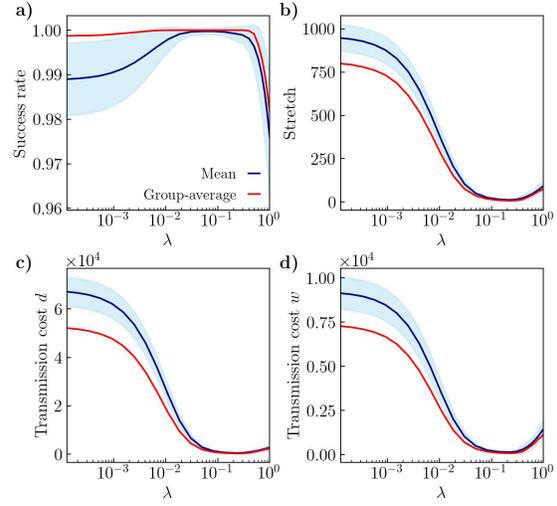

**SF 23**: Same as Fig. **SF 20** but for t=10000.

**SF 24**: Same as Fig. **SF 20** but for t=30000.

## F. Distribution of mean distances and λ-ratios

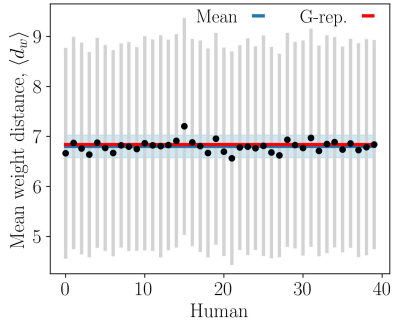 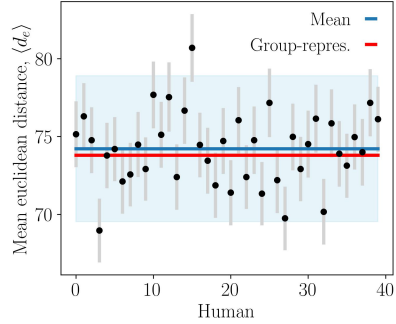 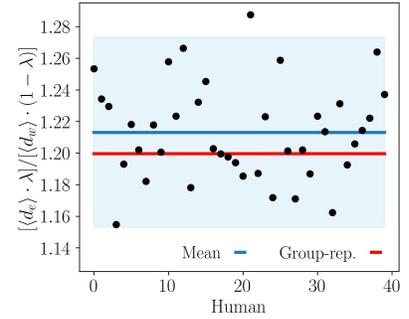

**SF 25**: Mean weight distance between pairs of connected nodes for each subject in the dataset (black dots) with its standard deviation (gray lines). The blue line represents the average mean weight distance of the cohort and the shaded blue area is $2\sigma$. The red line corresponds to the mean weight distance of the group-representative connectome.

**SF 26**: Mean Euclidean distance between all pairs of nodes for each subject in the dataset (black dots) with its standard deviation (gray lines). The blue line represents the average mean Euclidean distance of the cohort and the shaded blue area is $2\sigma$. The red line corresponds to the mean Euclidean distance of the group-representative connectome.

**SF 27**: Ratio between Euclidean distance and weight distance scaled by $\lambda = 0.1$ for each subject in the dataset (black dots). The blue line represents the average ratio of the cohort and the shaded blue area is $2\sigma$. The red line corresponds to the ratio of the group-representative connectome.

## G.   Navigation after weight-reshuffling

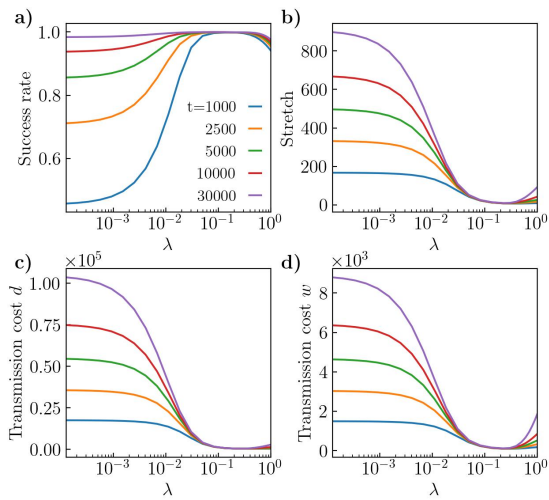

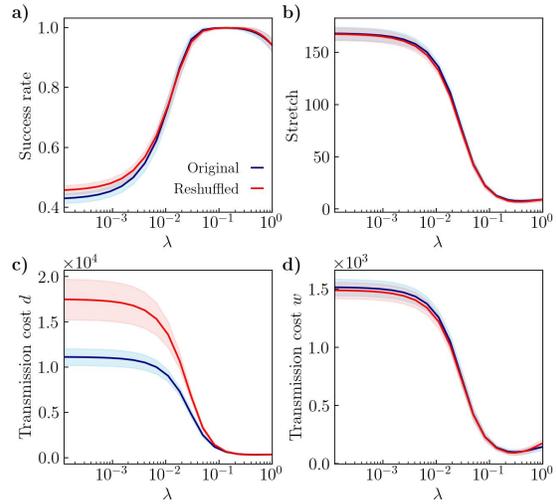

**SF 28**: Mean results across the 40 subjects of the UL Dataset after weight-reshuffling of **a)** success rate, **b)** stretch, **c)** Euclidean distance transmission cost and **d)** weight distance transmission cost. Each curve shows the outcome for a specific value of time-out.

**SF 29**: t=1000. Comparison of the mean results after the weight-reshuffling (in red) and the original mean results (in blue). The shadowed areas correspond to two times the standard deviation between subjects.

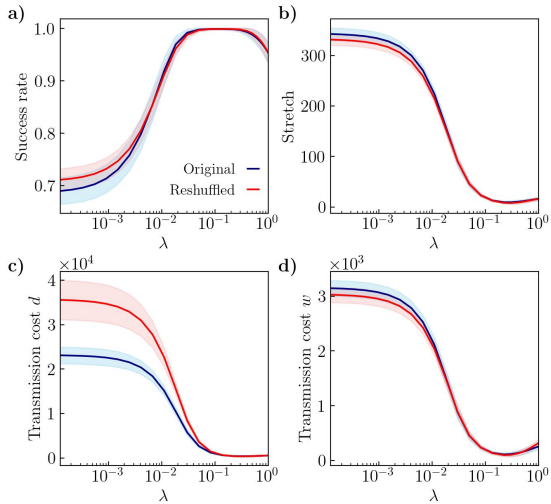

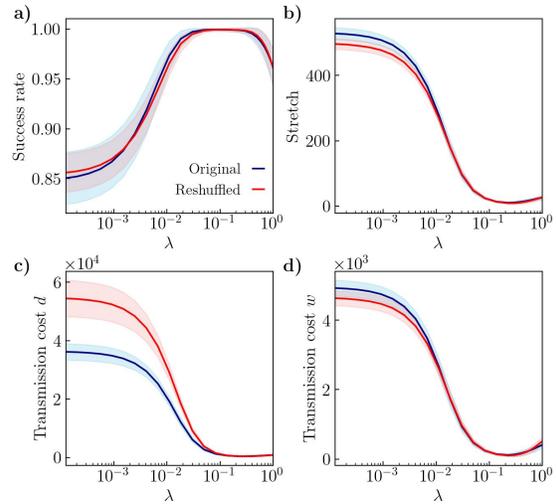

**SF 30**: Same as Fig. **SF 29** but for t=2500.

**SF 31**: Same as Fig. **SF 29** but for t=5000.

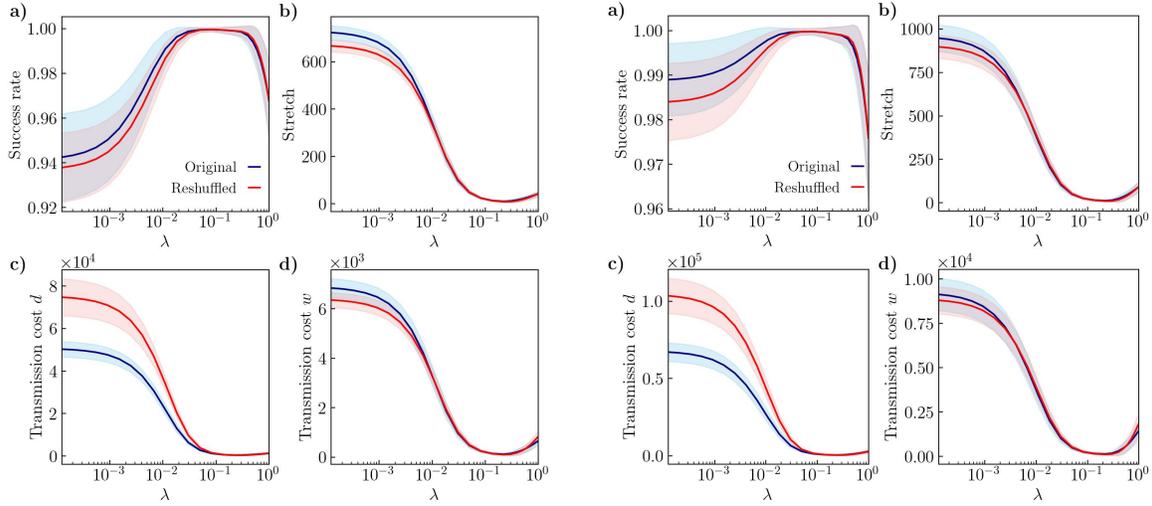

**SF 32**: Same as Fig. **SF 29** but for t=10000.       **SF 33**: Same as Fig. **SF 29** but for t=30000.

### H.   Errors in weight-reshuffling navigation for Human 34 and t=5000

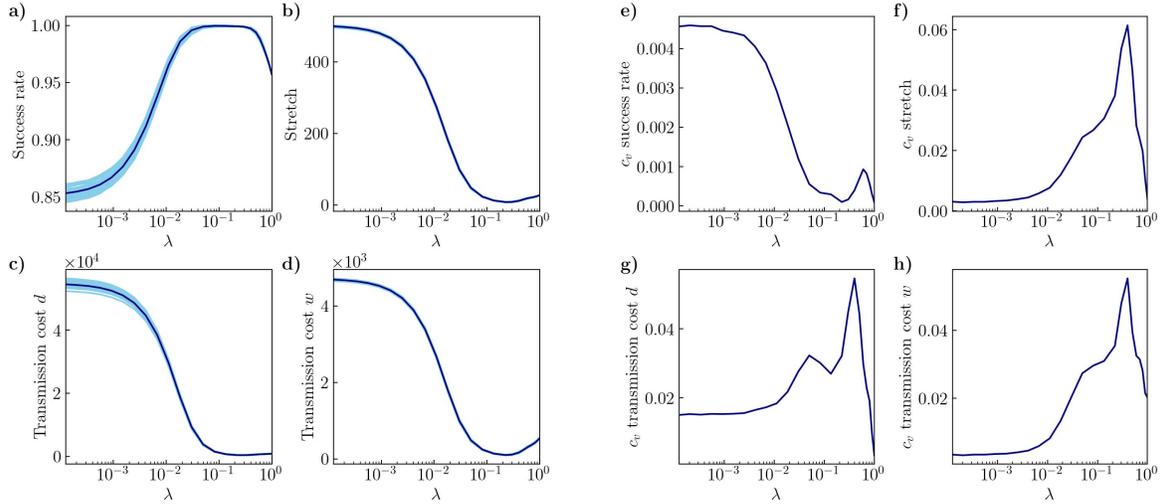

**SF 34**: **a-d)** 50 navigation realizations for subject Human 34 and time-out 5000 after weight-reshuffling. The mean of all realizations is represented in dark blue and each of the realizations in light blue. **e-h)** Coefficient of variation $c_v = \frac{\sigma}{\mu}$, where $\mu$ is the mean of all realizations and $\sigma$ is their standard deviation.

**I.    Navigation after position-reshuffling**

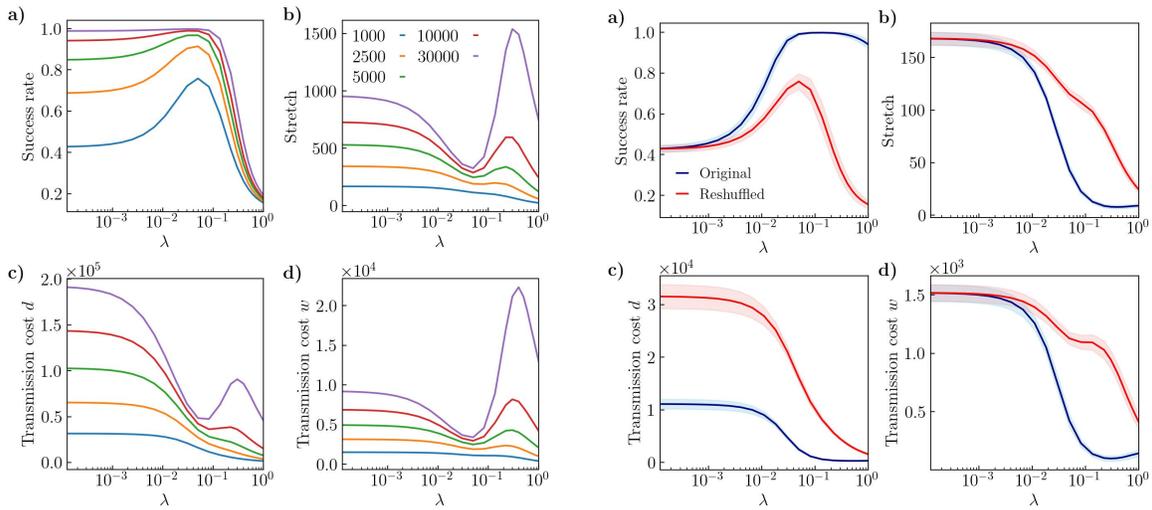

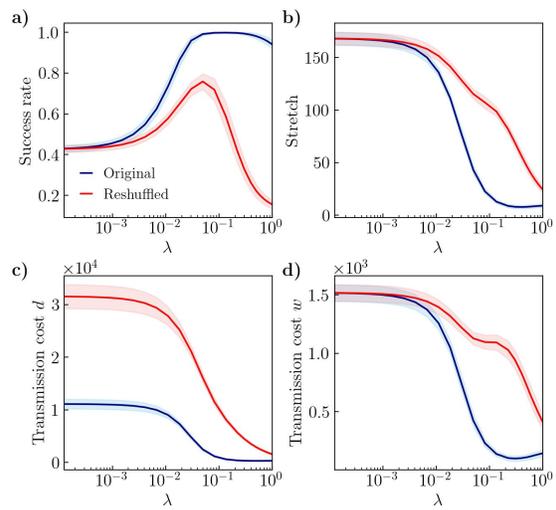

**SF 35**: Mean results across the 40 subjects of the UL Dataset after position-reshuffling of **a)** success rate, **b)** stretch, **c)** Euclidean distance transmission cost and **d)** weight distance transmission cost. Each curve shows the outcome for a specific value of time-out.

**SF 36**: t=1000. Comparison of the mean results after the position-reshuffling (in red) and the original mean results (in blue). The shadowed areas correspond to two times the standard deviation between subjects.

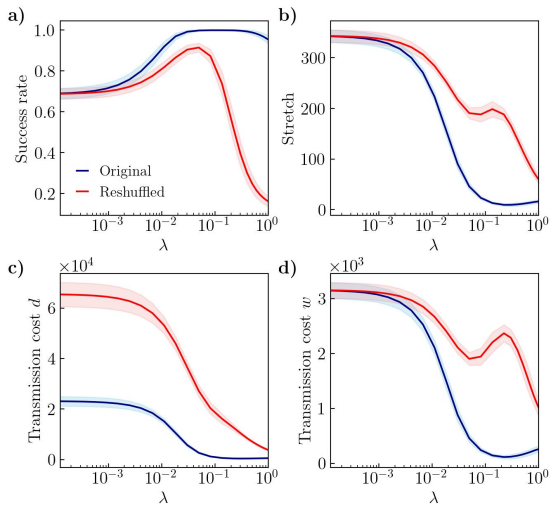

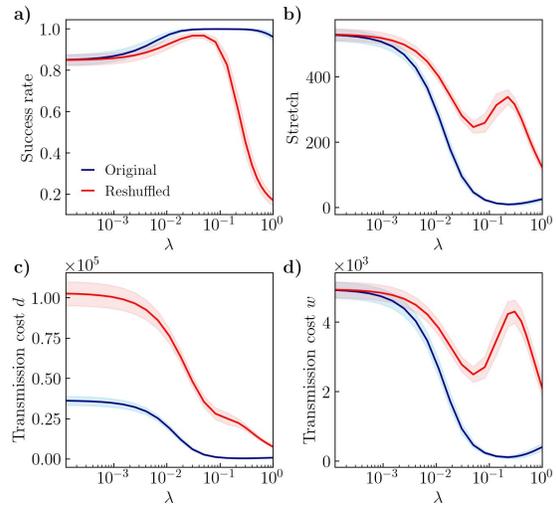

**SF 37**: Same as Fig. **SF 36** but for t=2500.

**SF 38**: Same as Fig. **SF 36** but for t=5000.

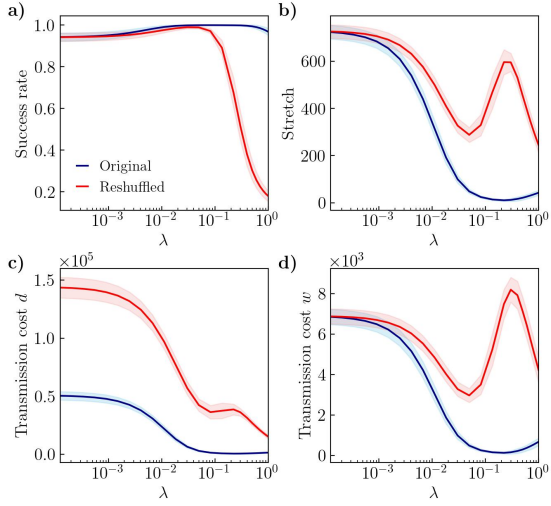

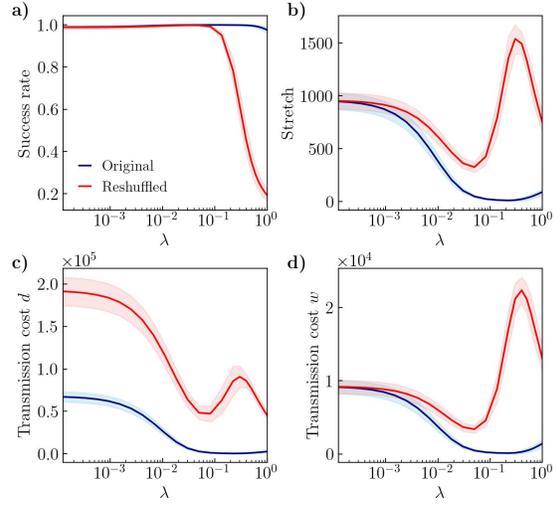

**SF 39**: Same as Fig. **SF 36** but for t=10000.

**SF 40**: Same as Fig. **SF 36** but for t=30000.

### J. Errors in position-reshuffling navigation for Human 34 and t=5000

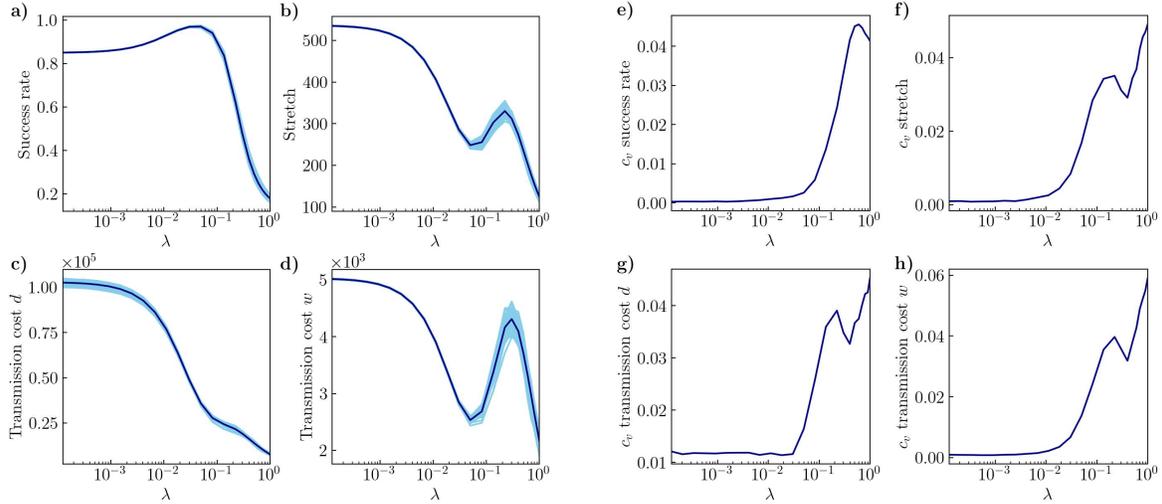

**SF 41**: **a-d)** 50 navigation realizations for subject Human 34 and time-out 5000 after position-reshuffling. The mean of all realizations is represented in dark blue and each of the realizations in light blue. **e-h)** Coefficient of variation $c_v = \frac{\sigma}{\mu}$, where $\mu$ is the mean of all realizations and $\sigma$ is their standard deviation.

## K. Attack on group-representative network for t=1000

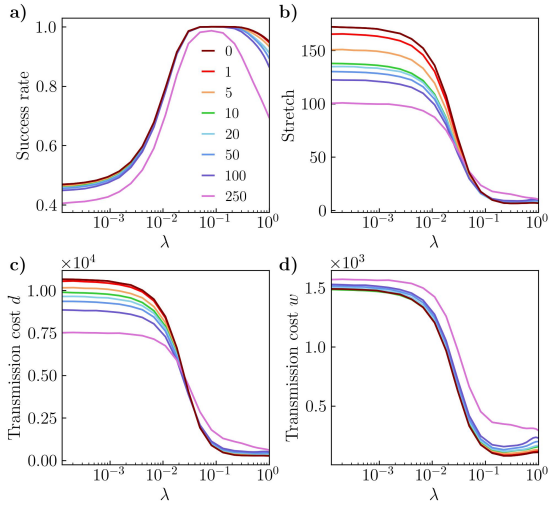

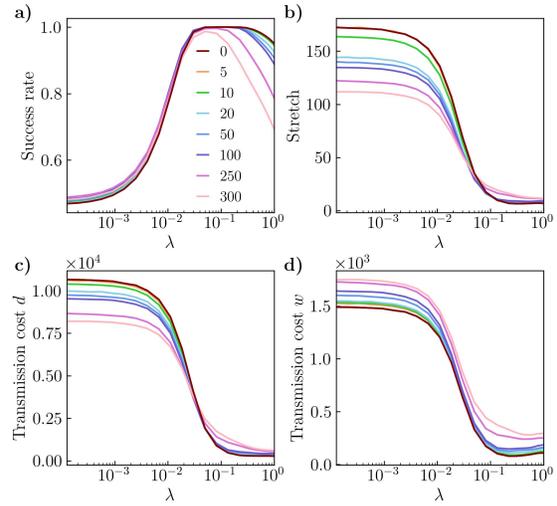

**SF 42**: Navigation results after attack on nodes with highest degree. Results of **a)** success rate, **b)** stretch, **c)** Euclidean distance transmission cost and **d)** weight distance transmission cost. The legend corresponds to the number of nodes deleted.

**SF 43**: Navigation results after attack on nodes with highest strength. The legend corresponds to the number of nodes deleted.

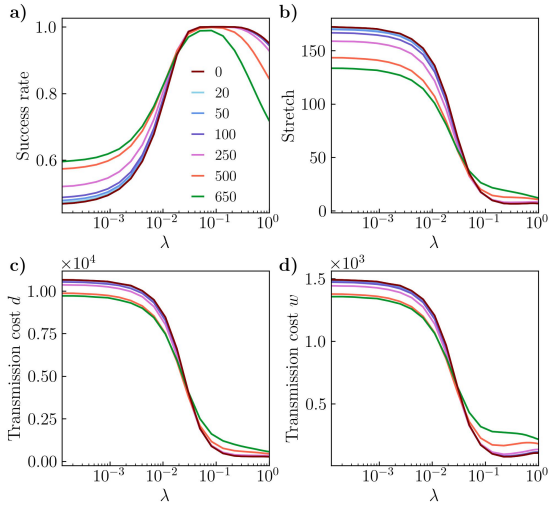

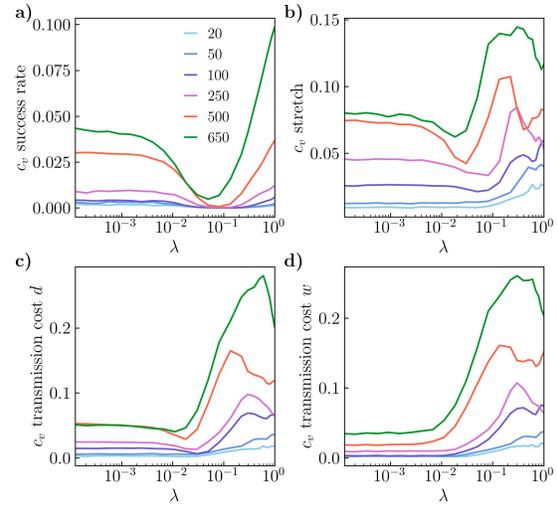

**SF 44**: Navigation results after random attack. Mean results of 10 realizations. The legend corresponds to the number of nodes deleted.

**SF 45**: Coefficient of variation $c_v = \frac{\sigma}{\mu}$, where $\mu$ is the mean of the realizations for the random attack and $\sigma$ is their standard deviation. The legend corresponds to the number of nodes deleted.

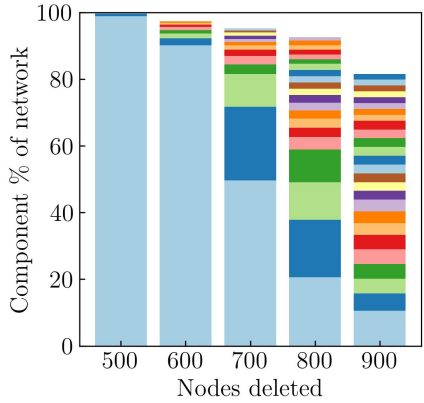

**SF 46**: Percentage of the group-representative network that each connected component represents, after the deletion of 500,600,700,800 and 900 nodes in decreasing order of degree. Each colored area corresponds to one component.

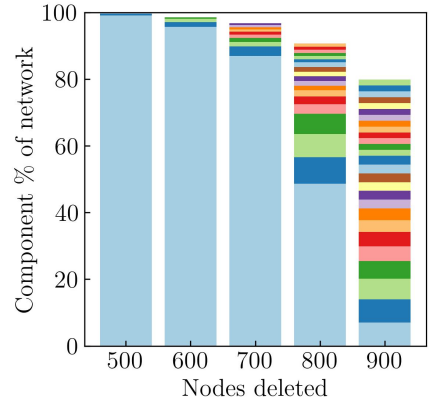

**SF 47**: Percentage of the group-representative network that each connected component represents, after the deletion of 500,600,700,800 and 900 nodes in decreasing order of strength. Each colored area corresponds to one component.

### L.  Navigation with homogeneous weights for t=5000

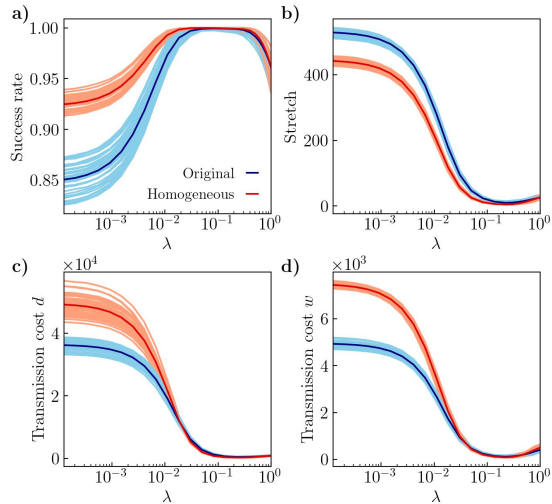

**SF 48**: t=5000. Results of **a)** success rate, **b)** stretch, **c)** Euclidean distance transmission cost and **d)** weight distance transmission cost. Original network: Sky-blue curves are the results for each subject and the dark-blue curve represents the mean of all subjects. Network with homogeneous weights (weight-distances between 5.5 and 7.5): Light-salmon curves are the results for each subject and the red curve represents the mean of all subjects.

## M. Navigation results with second framework

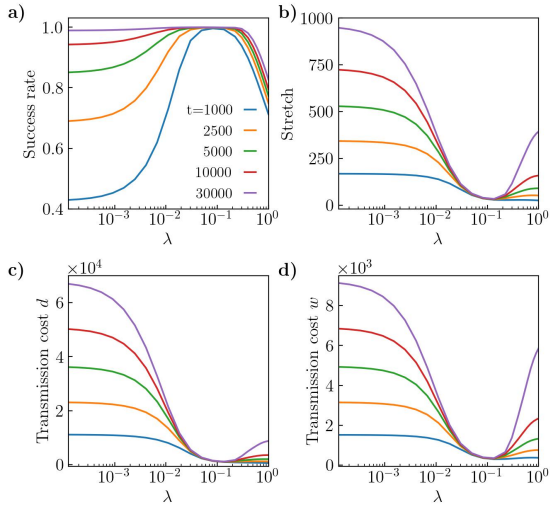

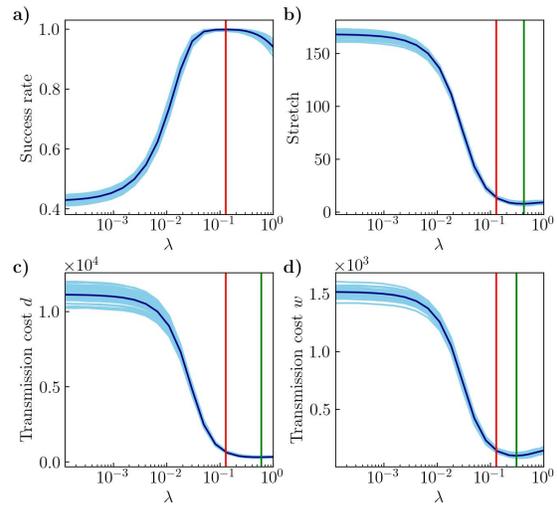

**SF 49**: Mean results across the 40 subjects of the UL Dataset of **a)** success rate, **b)** stretch, **c)** Euclidean distance transmission cost and **d)** weight distance transmission cost. Each curve shows the outcome for a specific value of time-out.

**SF 50**: t=1000. Sky-blue curves are the results for each subject. The dark-blue curve represents the mean. The $\lambda$ for the maximum value of success rate (**a**) is indicated using a vertical red line in all plots and in the plots of the stretch, Euclidean and weight distance transmission costs (**b,c,d**) the $\lambda$ for the minimum value of the respective magnitude is shown using a vertical green line.

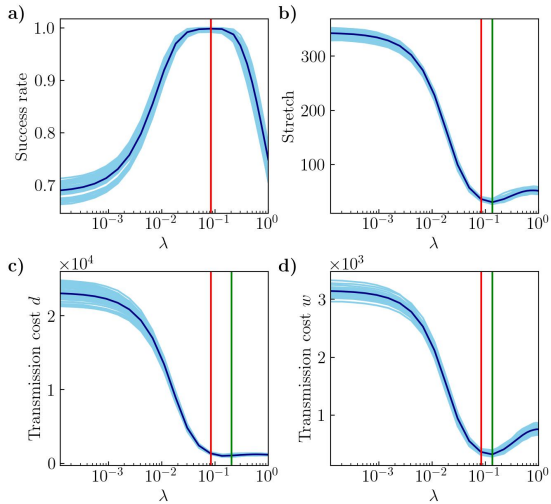

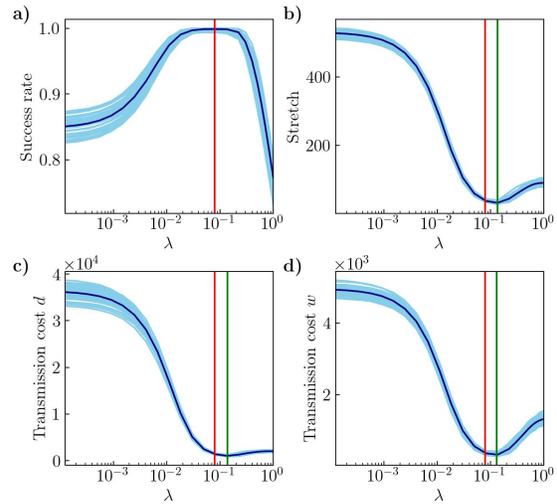

**SF 51**: Same as Fig. **SF 50** but for t=2500.

**SF 52**: Same as Fig. **SF 50** but for t=5000.

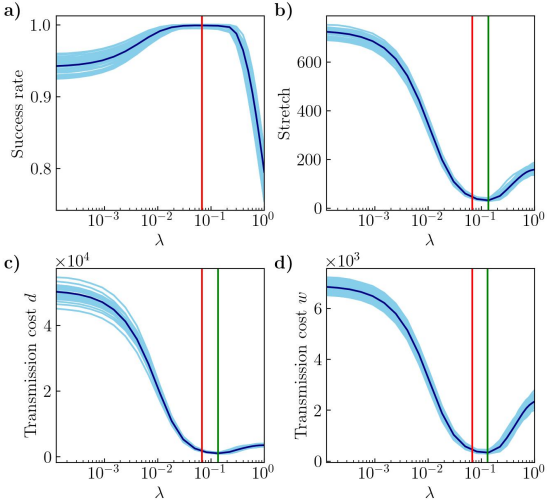 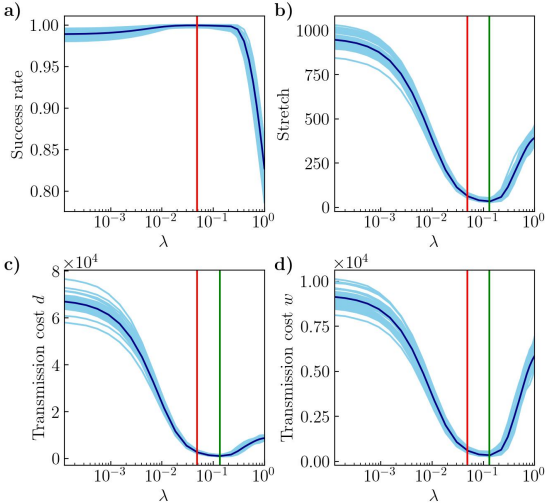

**SF 53**: Same as Fig. **SF 50** but for t=10000.   **SF 54**: Same as Fig. **SF 50** but for t=30000.

## II. RESULTS FOR HCP DATASET

### A. Network properties

| Human | Nodes | Edges | $\langle k \rangle$ | $k_{max}$ | $\langle s \rangle$ | $s_{max}$ | $\langle d_{nn} \rangle$ | $\langle d_{nn}^{max} \rangle$ | $\langle \lambda \rangle$ | Diameter |
|---|---|---|---|---|---|---|---|---|---|---|
| 0 | 1014 | 37910 | 74.773 | 529 | 0.093 | 0.408 | 40.213 | 97.596 | 2.107 | 4 |
| 1 | 1014 | 39210 | 77.337 | 602 | 0.081 | 0.411 | 42.020 | 103.189 | 2.071 | 4 |
| 2 | 1014 | 39609 | 78.124 | 566 | 0.102 | 0.502 | 39.477 | 95.695 | 2.093 | 4 |
| 3 | 1014 | 40309 | 79.505 | 603 | 0.092 | 0.445 | 41.703 | 104.094 | 2.053 | 4 |
| 4 | 1014 | 43276 | 85.357 | 678 | 0.103 | 0.459 | 40.709 | 99.200 | 2.009 | 4 |
| 5 | 1014 | 40165 | 79.221 | 616 | 0.077 | 0.349 | 44.284 | 100.322 | 2.067 | 4 |
| 6 | 1014 | 39638 | 78.181 | 605 | 0.099 | 0.482 | 40.290 | 98.492 | 2.049 | 4 |
| 7 | 1014 | 44089 | 86.961 | 645 | 0.098 | 0.467 | 40.809 | 99.934 | 2.031 | 4 |
| 8 | 1014 | 36557 | 72.105 | 539 | 0.124 | 0.666 | 35.970 | 88.843 | 2.129 | 5 |
| 9 | 1014 | 38249 | 75.442 | 536 | 0.115 | 0.566 | 39.968 | 96.122 | 2.069 | 4 |
| 10 | 1014 | 39578 | 78.063 | 574 | 0.067 | 0.357 | 43.026 | 105.373 | 2.094 | 4 |
| 11 | 1013 | 34436 | 67.988 | 475 | 0.093 | 0.390 | 38.590 | 94.119 | 2.152 | 5 |
| 12 | 1014 | 39631 | 78.168 | 535 | 0.109 | 0.419 | 39.412 | 96.746 | 2.089 | 4 |
| 13 | 1014 | 40358 | 79.602 | 630 | 0.078 | 0.379 | 41.729 | 102.733 | 2.048 | 4 |
| 14 | 1014 | 40821 | 80.515 | 581 | 0.091 | 0.357 | 39.831 | 96.191 | 2.072 | 4 |
| 15 | 1014 | 37896 | 74.746 | 589 | 0.075 | 0.335 | 43.490 | 106.477 | 2.052 | 4 |
| 16 | 1014 | 39589 | 78.085 | 567 | 0.075 | 0.357 | 43.146 | 105.475 | 2.067 | 4 |
| 17 | 1014 | 38009 | 74.968 | 555 | 0.107 | 0.505 | 39.038 | 94.241 | 2.105 | 5 |
| 18 | 1014 | 46070 | 90.868 | 604 | 0.078 | 0.347 | 43.319 | 106.259 | 2.007 | 4 |
| 19 | 1014 | 41676 | 82.201 | 610 | 0.097 | 0.412 | 40.406 | 100.044 | 2.040 | 4 |
| 20 | 1013 | 41727 | 82.383 | 592 | 0.116 | 0.504 | 38.529 | 94.514 | 2.057 | 4 |
| 21 | 1014 | 36659 | 72.306 | 579 | 0.106 | 0.472 | 39.467 | 96.305 | 2.089 | 4 |
| 22 | 1014 | 38855 | 76.637 | 598 | 0.062 | 0.263 | 45.439 | 110.573 | 2.076 | 4 |
| 23 | 1014 | 37235 | 73.442 | 609 | 0.078 | 0.331 | 42.697 | 102.966 | 2.077 | 4 |
| 24 | 1014 | 38407 | 75.753 | 559 | 0.085 | 0.413 | 41.131 | 101.004 | 2.069 | 4 |
| 25 | 1014 | 43631 | 86.057 | 623 | 0.097 | 0.438 | 40.056 | 99.229 | 2.024 | 4 |
| 26 | 1014 | 42351 | 83.533 | 581 | 0.100 | 0.428 | 40.746 | 98.902 | 2.060 | 4 |
| 27 | 1014 | 37875 | 74.704 | 527 | 0.080 | 0.364 | 41.813 | 101.904 | 2.103 | 4 |
| 28 | 1014 | 35737 | 70.487 | 558 | 0.092 | 0.447 | 40.679 | 99.079 | 2.100 | 4 |
| 29 | 1014 | 39125 | 77.170 | 542 | 0.097 | 0.458 | 38.723 | 93.705 | 2.097 | 4 |
| 30 | 1014 | 41041 | 80.949 | 586 | 0.104 | 0.484 | 40.165 | 96.655 | 2.068 | 4 |
| 31 | 1014 | 39813 | 78.527 | 650 | 0.079 | 0.381 | 43.223 | 105.802 | 2.059 | 4 |
| 32 | 1013 | 39161 | 77.317 | 571 | 0.067 | 0.326 | 43.563 | 105.526 | 2.086 | 4 |
| 33 | 1014 | 40400 | 79.684 | 682 | 0.080 | 0.355 | 44.100 | 108.783 | 2.056 | 4 |
| 34 | 1014 | 44274 | 87.325 | 643 | 0.094 | 0.451 | 40.880 | 100.170 | 2.010 | 4 |
| 35 | 1014 | 46025 | 90.779 | 632 | 0.090 | 0.451 | 43.121 | 105.236 | 2.014 | 4 |
| 36 | 1014 | 37237 | 73.446 | 578 | 0.063 | 0.247 | 44.849 | 107.836 | 2.094 | 4 |
| 37 | 1014 | 45512 | 89.767 | 618 | 0.077 | 0.362 | 43.105 | 106.535 | 2.011 | 4 |
| 38 | 1014 | 40209 | 79.308 | 647 | 0.080 | 0.378 | 44.132 | 107.810 | 2.055 | 4 |
| 39 | 1014 | 34947 | 68.929 | 623 | 0.078 | 0.378 | 40.932 | 99.532 | 2.090 | 4 |
| 40 | 1014 | 38384 | 75.708 | 592 | 0.103 | 0.534 | 39.196 | 96.133 | 2.076 | 4 |
| 41 | 1014 | 35161 | 69.351 | 506 | 0.091 | 0.466 | 40.540 | 99.649 | 2.115 | 4 |
| 42 | 1014 | 40731 | 80.337 | 591 | 0.098 | 0.462 | 41.340 | 99.343 | 2.068 | 4 |
| 43 | 1014 | 37845 | 74.645 | 519 | 0.084 | 0.375 | 39.877 | 98.111 | 2.107 | 4 |
| Average | 1013.932 | 39759.5 | 78.426 | 587.386 | 0.090 | 0.418 | 41.267 | 100.806 | 2.070 | 4.068 |
| Maximum | 1014 | 46070 | 90.868 | 682 | 0.124 | 0.666 | 45.439 | 110.573 | 2.152 | 5 |
| Minimum | 1013 | 34436 | 67.988 | 475 | 0.062 | 0.247 | 35.970 | 88.843 | 2.007 | 4 |
| Group-representative | 1014 | 38771 | 76.471 | 717 | 0.097 | 0.424 | 34.927 | 86.683 | 2.046 | 5 |

**ST 2**: Number of nodes, number of edges, average degree, maximum degree, average strength, maximum strength, mean average distance to nearest neighbors, mean maximum distance to nearest neighbors, average shortest path and diameter of the network for all connectomes in the HCP Dataset. The last four rows correspond to the mean between all subjects, the maximum of each of the columns, the minimum and the results for the group-representative connectome.

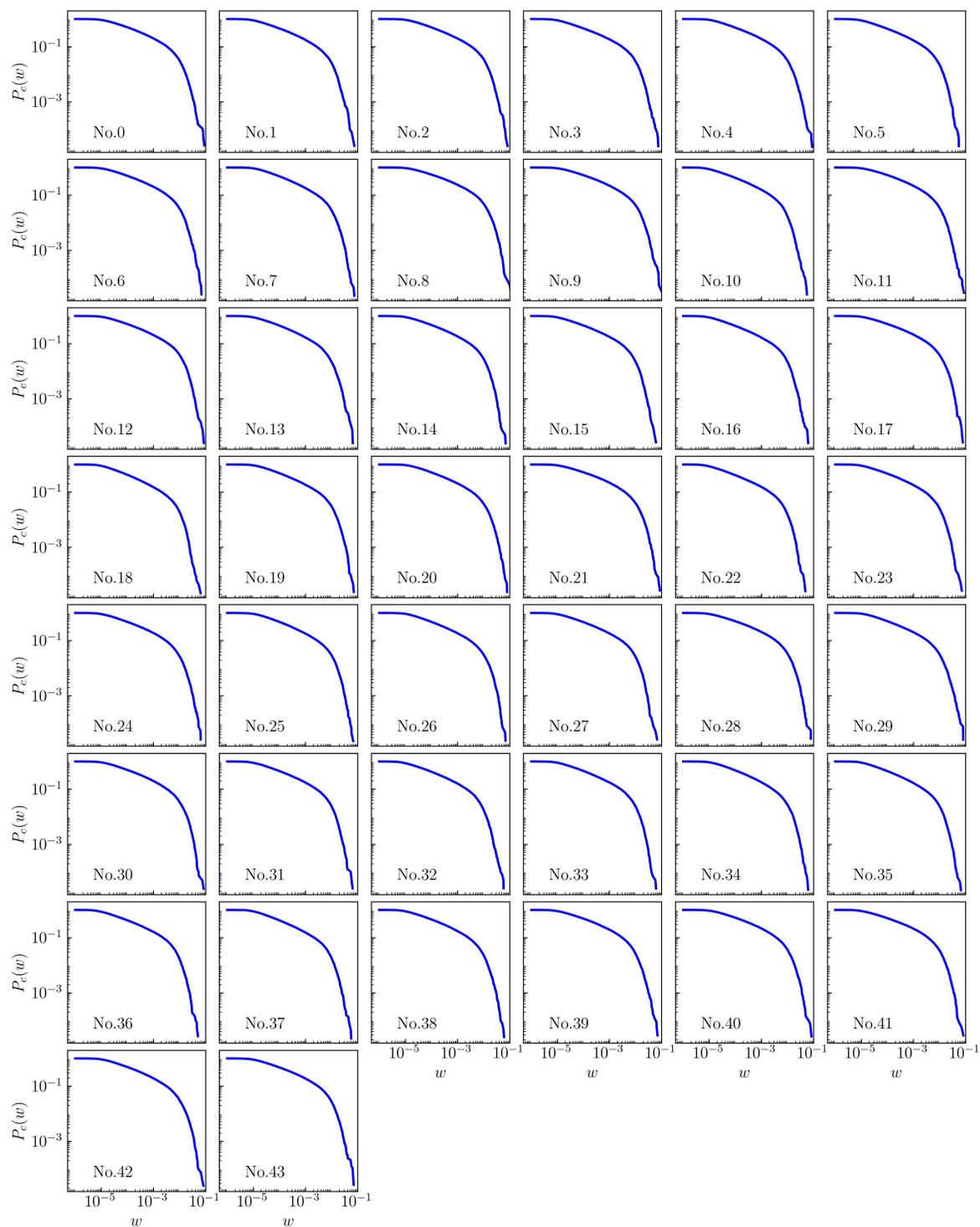

**SF 55**: Complementary cumulative weight distribution $P_c(w)$.

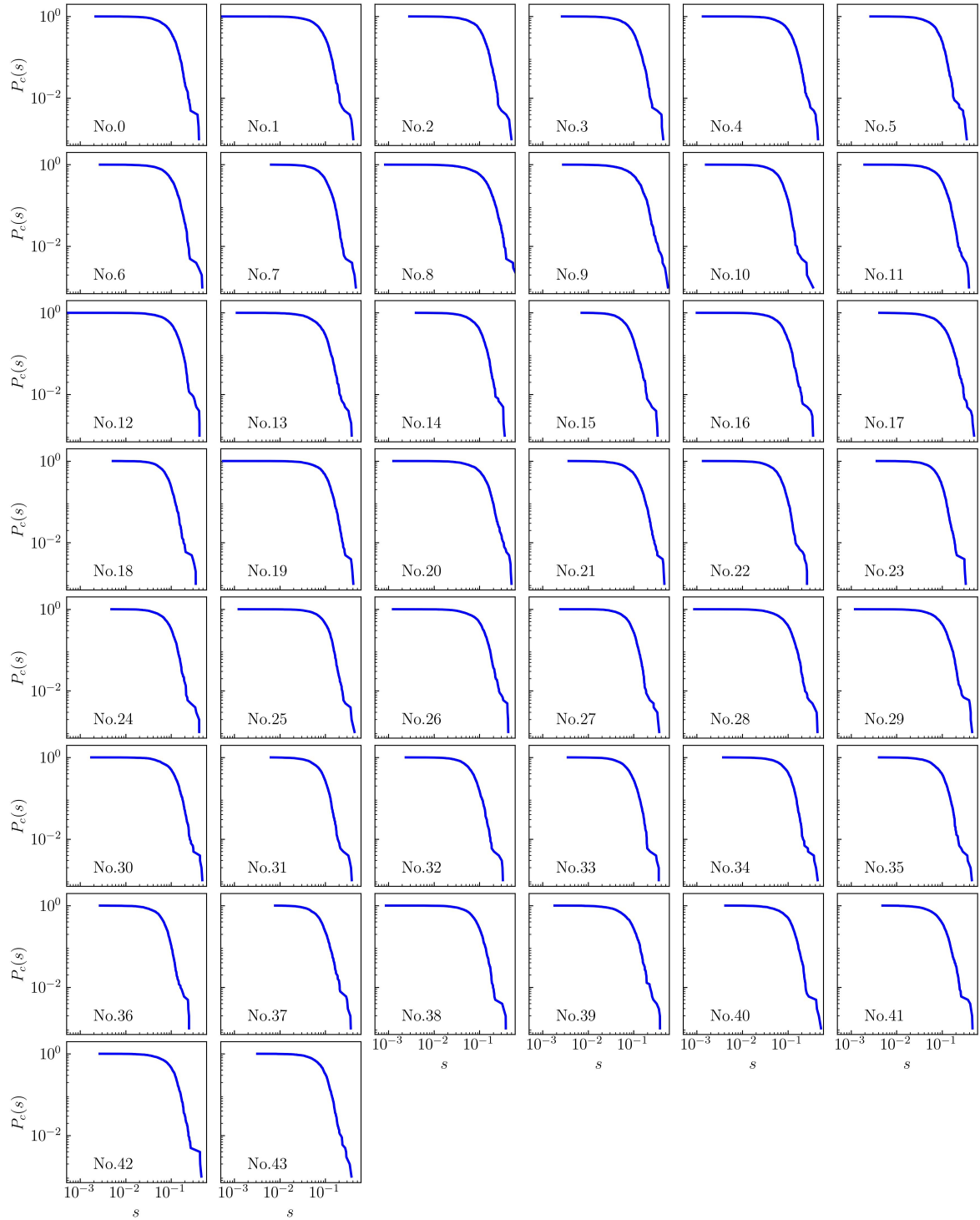

**SF 56**: Complementary cumulative strength distribution $P_c(s)$.

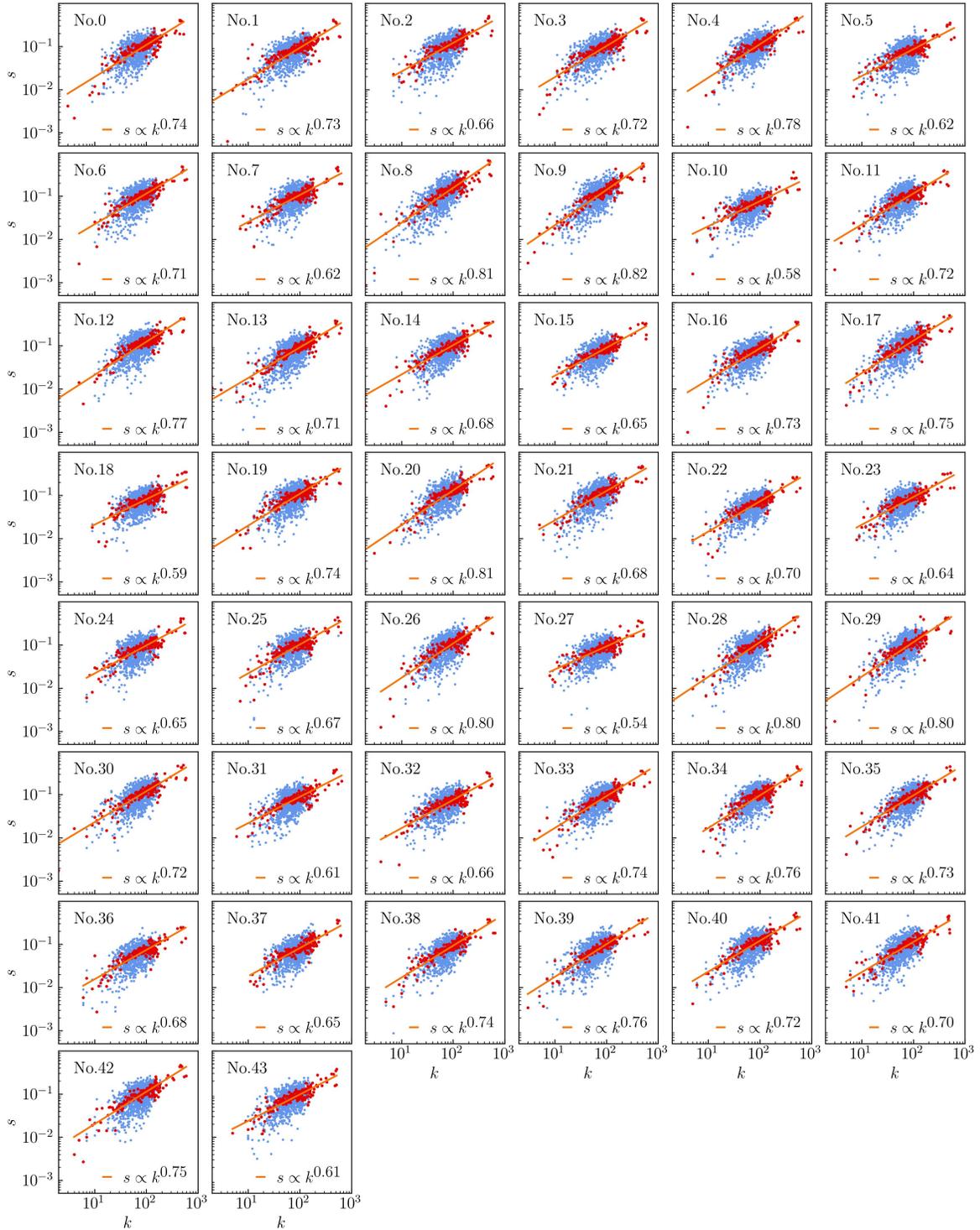

**SF 57**: Degree-strength correlation. Each of the blue dots corresponds to one node of the connectome. The red dots are the average values of strength for each degree. The orange curve is the fit of the averages. Average exponent: $\mu = 0.71 \pm 0.07$.

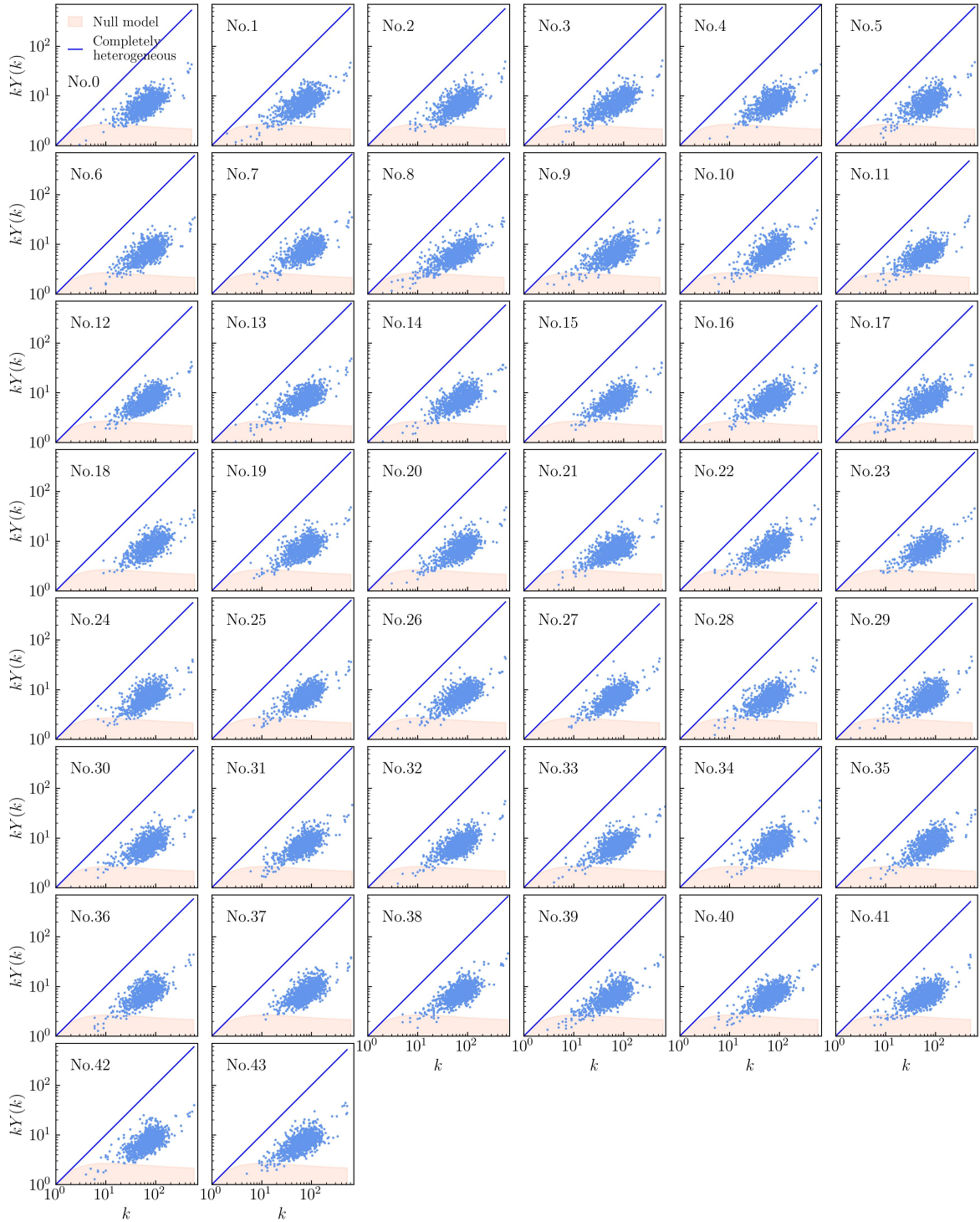

**SF 58**: Disparity measure. Each of the blue dots corresponds to one node of the connectome. The orange area corresponds to the average $+2\sigma$ given by the null model explained in Methods. The blue line represents the completely heterogeneous scenario.

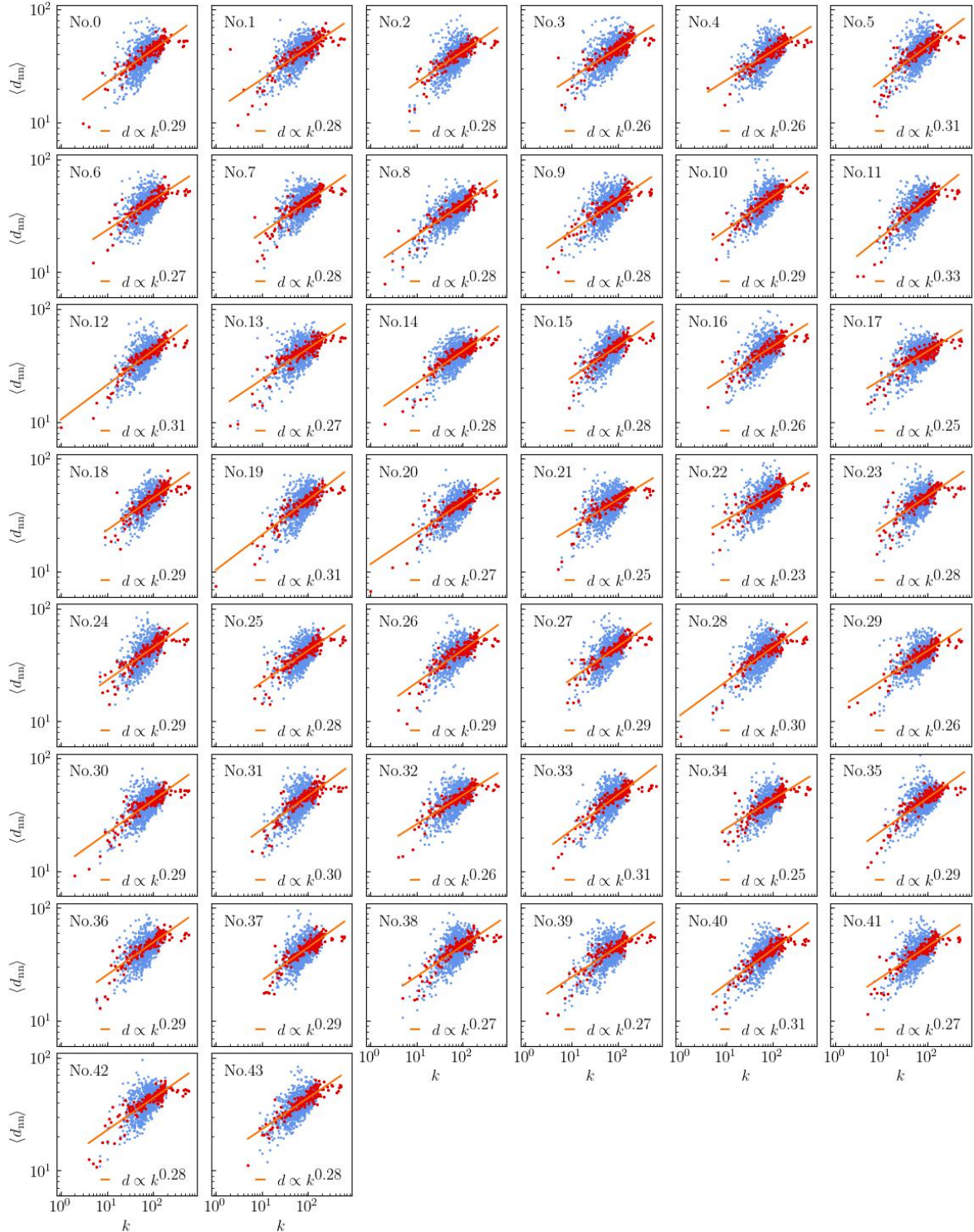

**SF 59**: Average Euclidean distance to nearest neighbours with respect to degree. Each of the blue dots corresponds to one node of the connectome. The red dots are the average values of Euclidean distance for each degree. The orange curve is the fit of the averages. Average exponent: $\mu = 0.281 \pm 0.019$.

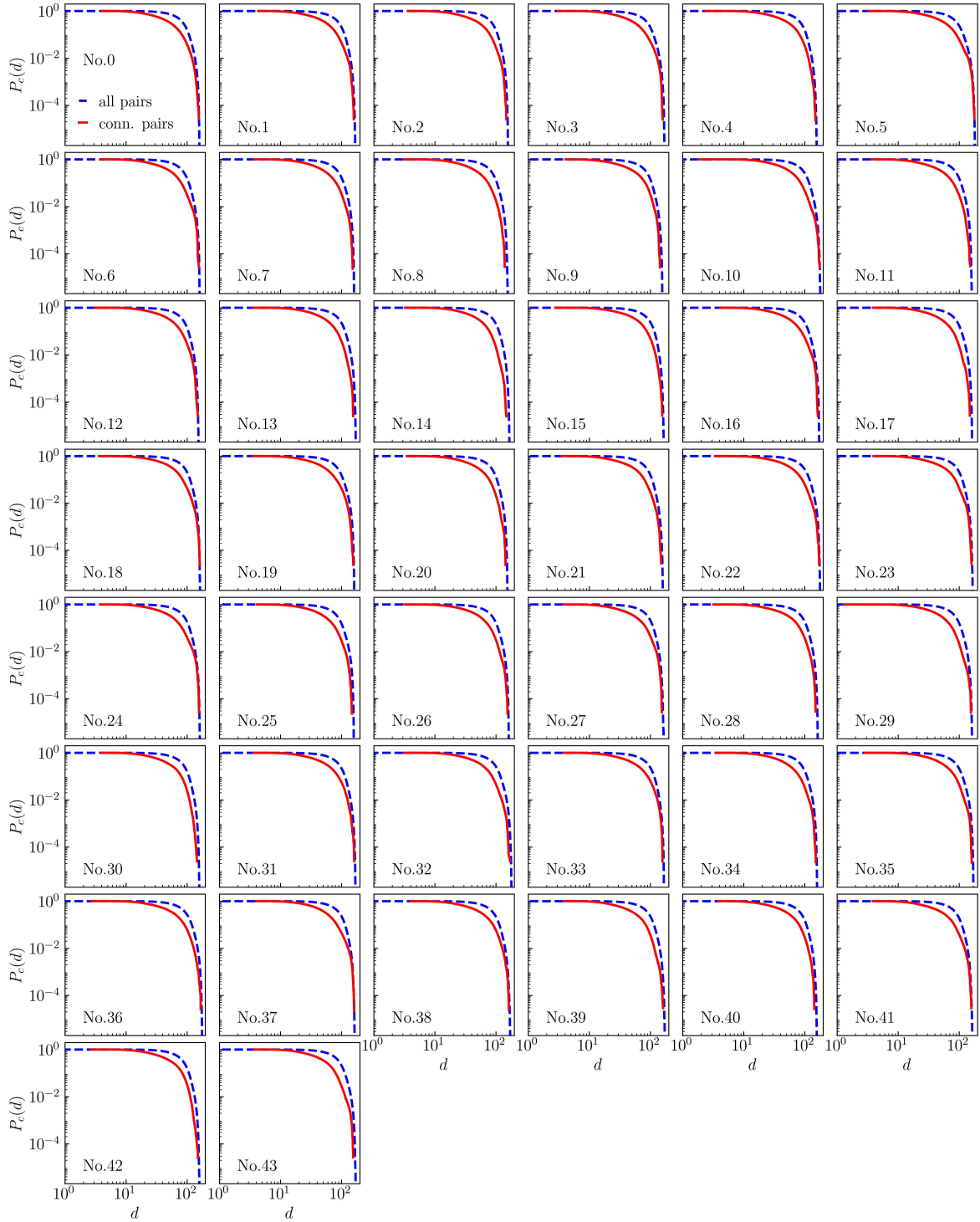

**SF 60**: Complementary cumulative Euclidean distance distribution, $P_c(d)$. The dashed blue curve corresponds to the distribution of distances between all pairs in the network and the red curve to the distances between connected pairs.

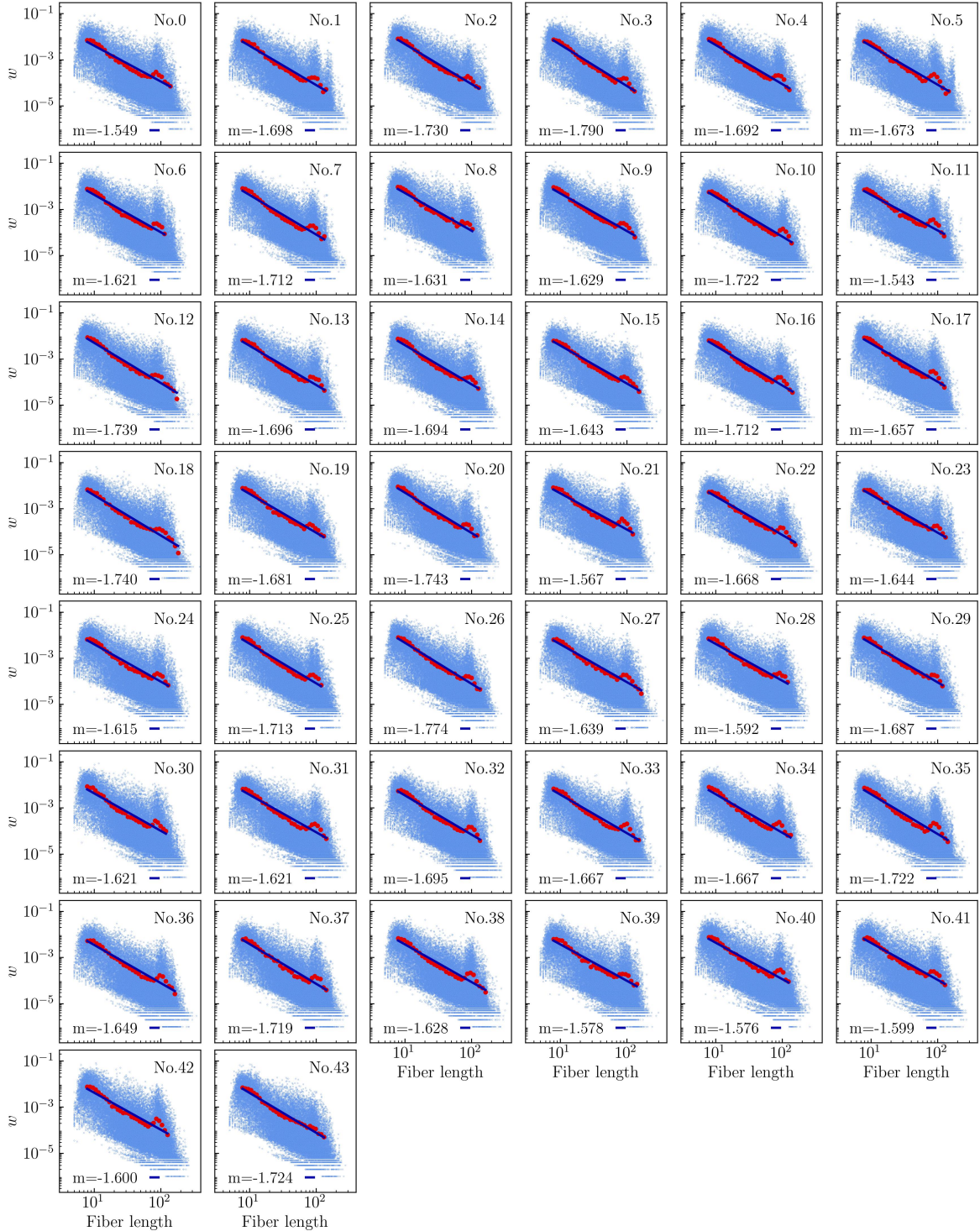

**SF 61**: Relatioship between weight and fiber length. The red dots correspond to the average weight for each of the 30 logarithmically-spaced fiber length bins ranging from flength$_{min}$×1.5 to flength$_{max}$×0.5. The fit of the averaged values is depicted in blue and it has the form $\ln(w) = m \ln(\text{flength}) + n$, with $\langle m \rangle = -1.67 \pm 0.06$.

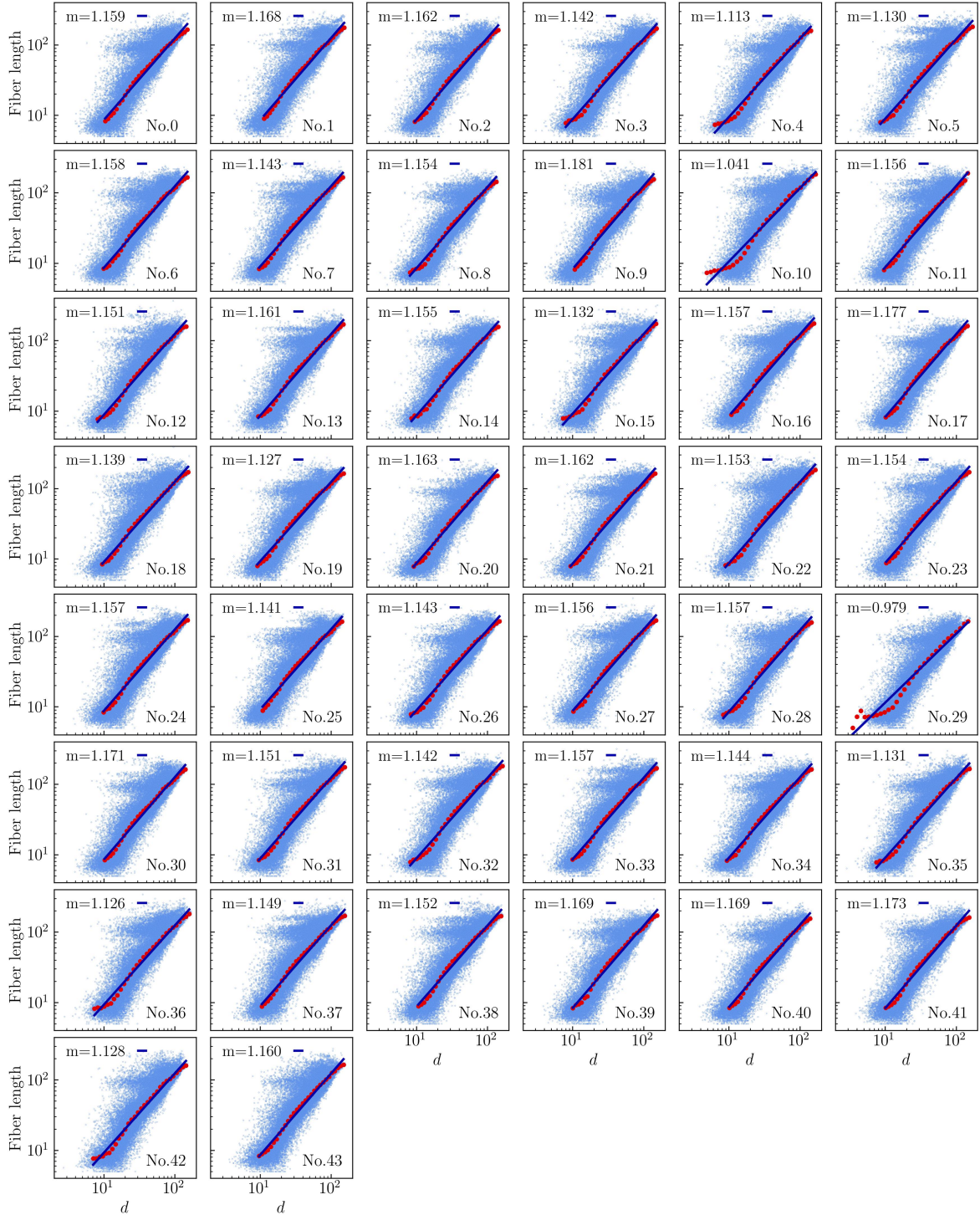

**SF 62**: Relatioship between fiber length and Euclidean distance. The red dots correspond to the average fiber length in each of the 30 logarithmically-spaced Euclidean distance bins ranging from $d_{\min} \times 1.5$ to $d_{\max} \times 0.5$. The fit of the averaged values is depicted in blue and it has the form $\ln(\text{flength}) = m \ln(d) + n$, with $\langle m \rangle = 1.15 \pm 0.04$.

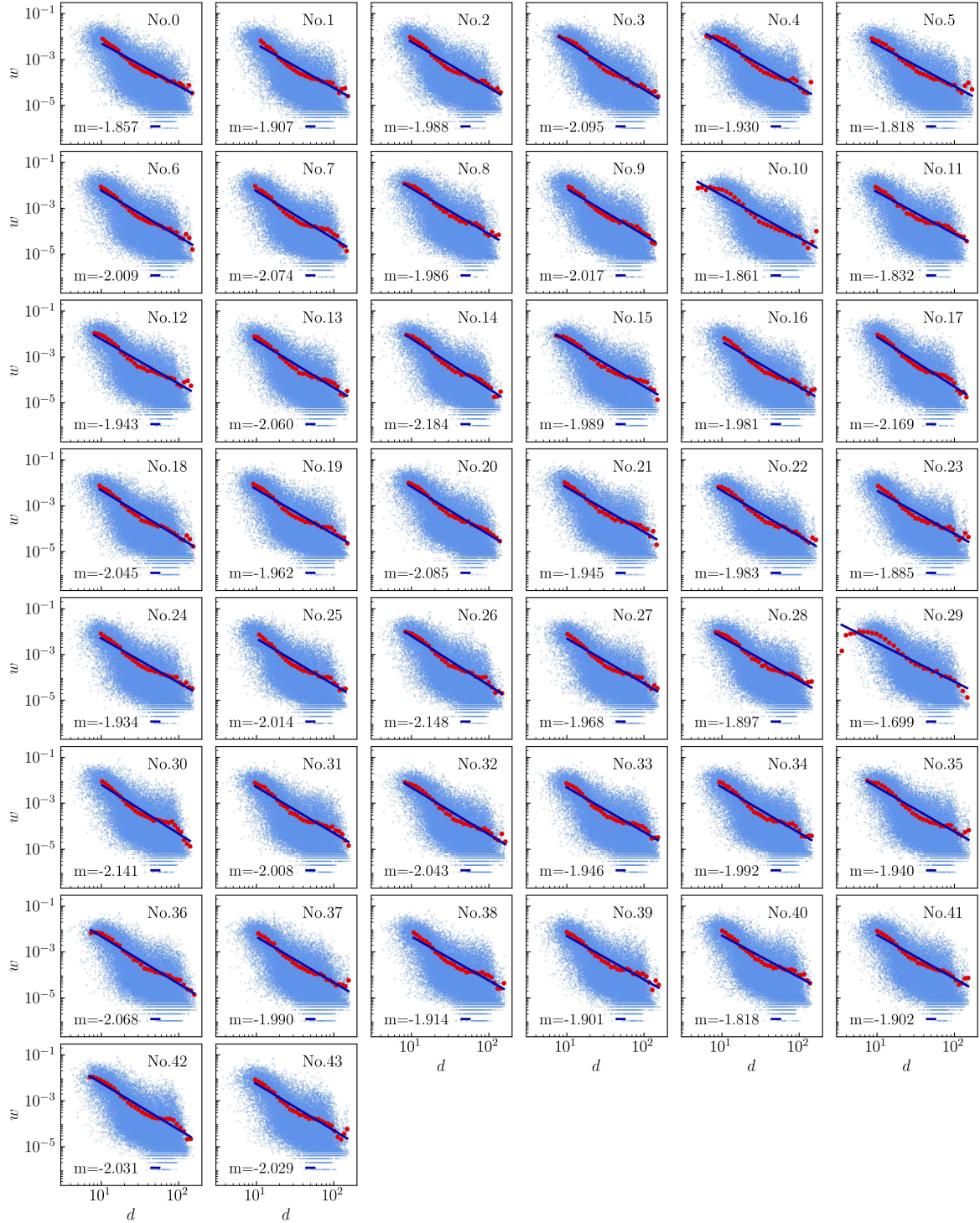

**SF 63**: Relatioship between weight and Euclidean distance. The red dots correspond to the average weight for each of the 30 logarithmically-spaced Euclidean distance bins ranging from $d_{\min} \times 2.5$ to $d_{\max}$. The fit of the averaged values is depicted in blue and it has the form $\ln(w) = m \ln(d) + n$, with $\langle m \rangle = -2.0 \pm 0.1$.

## B. Navigation results

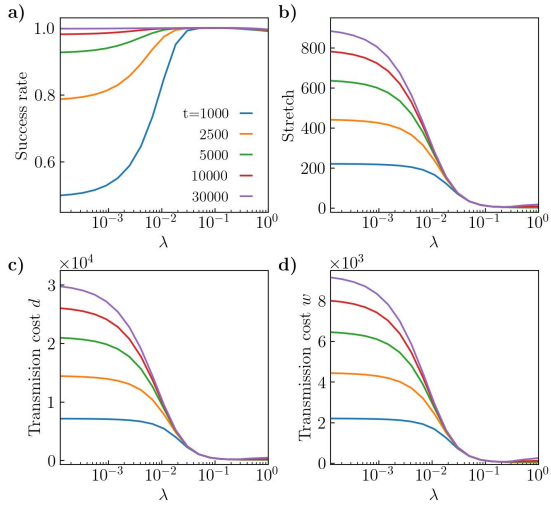

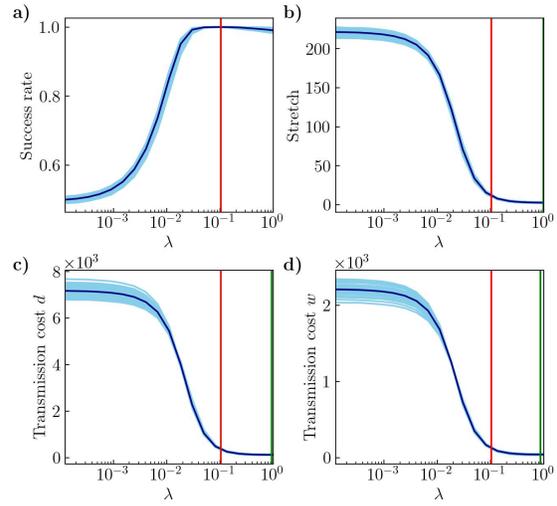

**SF 64**: Mean results across the 40 subjects of the HCP Dataset of **a)** success rate, **b)** stretch, **c)** Euclidean distance transmission cost and **d)** weight distance transmission cost. Each curve shows the outcome for a specific value of time-out.

**SF 65**: t=1000. Sky-blue curves are the results for each subject. The dark-blue curve represents the mean. The $\lambda$ for the maximum value of success rate **(a)** is indicated using a vertical red line in all plots, and in the plots of the stretch, Euclidean and weight distance transmission costs **(b,c,d)** the $\lambda$ for the minimum value of the respective magnitude is shown using a vertical green line.

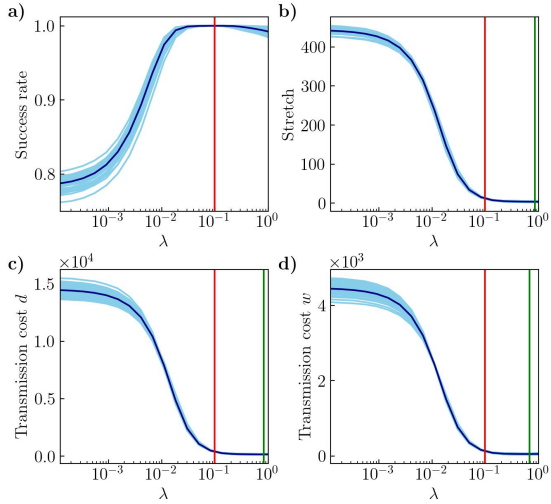

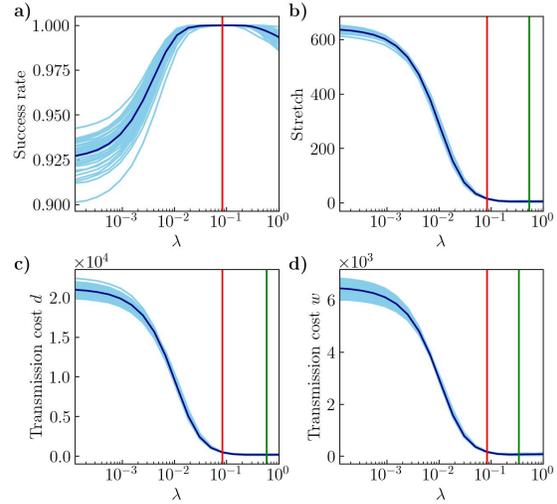

**SF 66**: Same as Fig. **SF 65** but for t=2500.

**SF 67**: Same as Fig. **SF 65** but for t=5000.

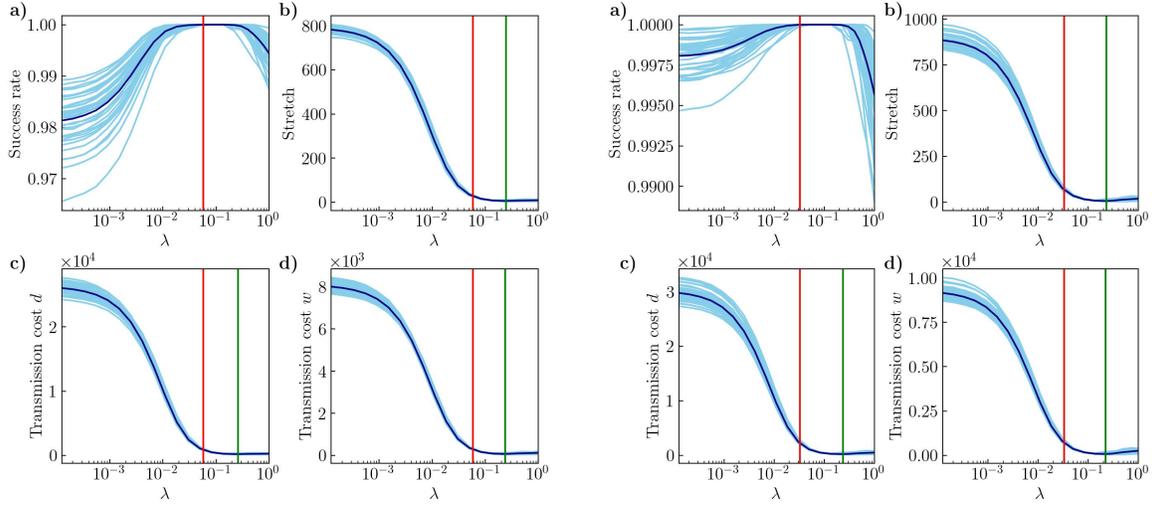

**SF 68**: Same as Fig. **SF 65** but for t=10000.  **SF 69**: Same as Fig. **SF 65** but for t=30000.

### C.  Group-representative connectome properties

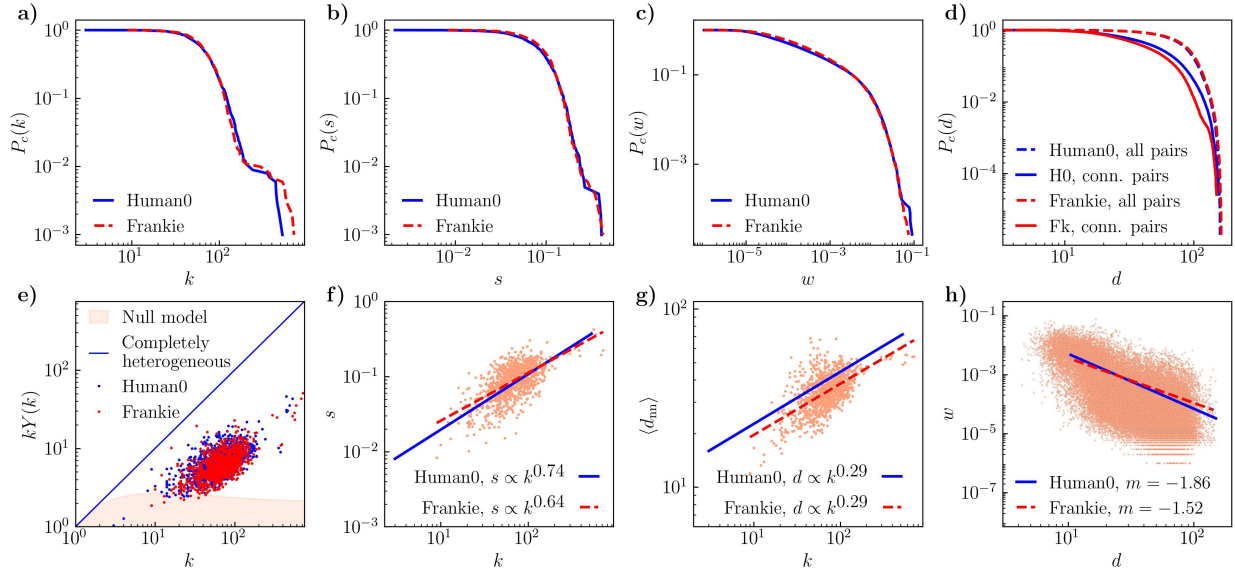

**SF 70**: Comparison of network properties between the group-representative connectome, referred as Frankie, and Human 0. **a)** Complementary cumulative degree distribution $P_c(k)$. **b)** Complementary cumulative strength distribution $P_c(s)$. **c)** Complementary cumulative weight distribution $P_c(w)$. **d)** Complementary cumulative Euclidean distance distribution $P_c(d)$. **e)** Disparity of weights. **f)** Degree-strength correlation. **g)** Average Euclidean distance to nearest neighbours vs degree. **h)** Relatioship between weight and Euclidean distance. In all plots the blue curves show the results for Human 0 and the red ones for the group-representative connectome. The orange dots in **f)**, **g)** and **h)** represent the results for each of the nodes of the group-representative connectome.

### D. Group-representative navigation

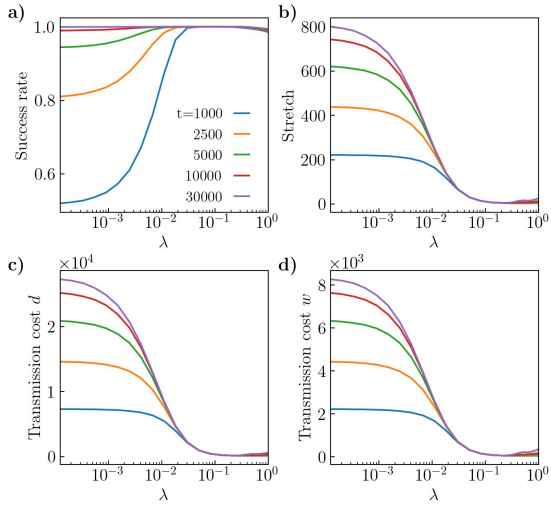

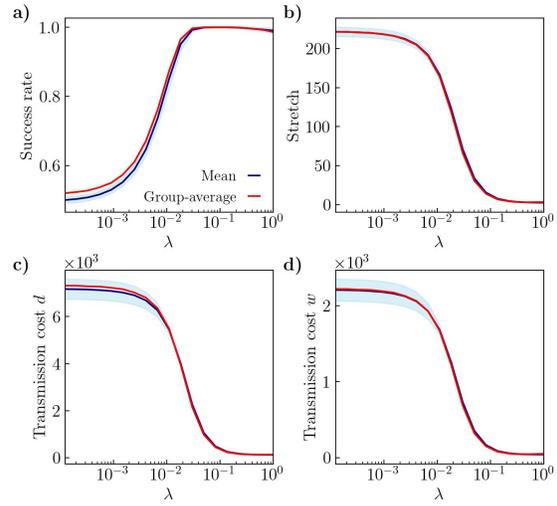

**SF 71**: **a)** Success rate, **b)** stretch, **c)** Euclidean distance transmission cost and **d)** weight distance transmission cost for the group-representative connectome of the HCP Dataset. Each curve shows the outcome for a specific value of time-out.

**SF 72**: t=1000. Comparison of the results for the group-representative (in red), here referred as group-average, and the mean results of all subjects (in blue). The blue-shadowed area corresponds to two times the standard deviation between subjects.

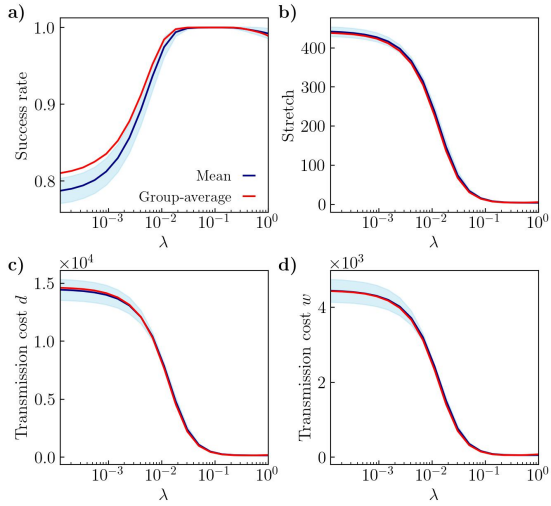

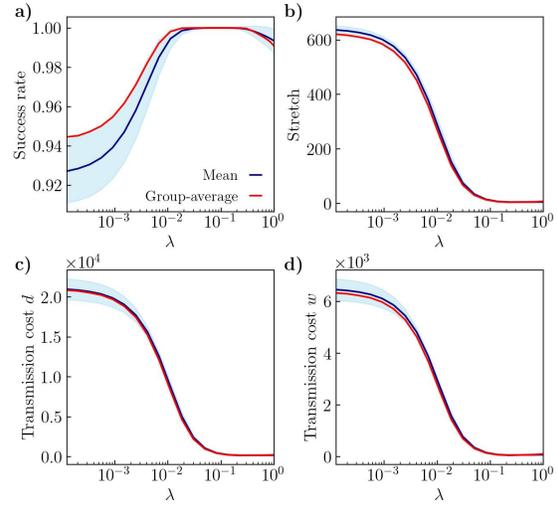

**SF 73**: Same as Fig. **SF 72** but for t=2500.

**SF 74**: Same as Fig. **SF 72** but for t=5000.

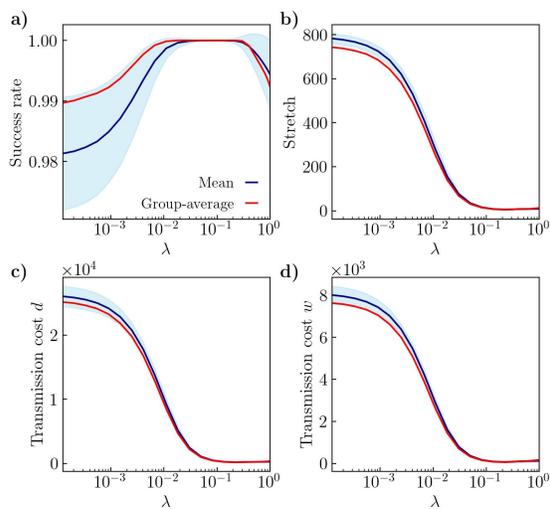

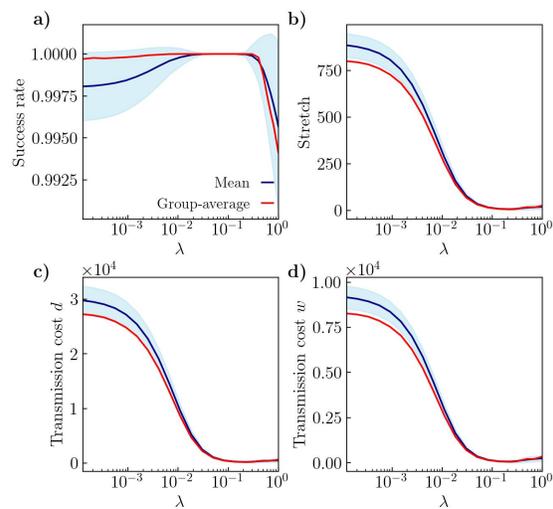

**SF 75**: Same as Fig. **SF 72** but for t=10000.

**SF 76**: Same as Fig. **SF 72** but for t=30000.


	
\end{document}